\documentclass[epj]{svjour}
\usepackage{latexsym}
\usepackage{graphics}
\begin{document}
\title{$\eta$-nucleus interaction from the $d+d$ reaction around the
$\eta$ production threshold}
\author{N. Ikeno\inst{1} \and H. Nagahiro\inst{2,3}
\and D. Jido\inst{4} \and S. Hirenzaki\inst{2} 
}                     
%
%
\institute{
Department of Life and Environmental Agricultural Sciences,
Tottori University, Tottori
680-8551, Japan
\and Department of Physics, Nara Women's University, Nara 630-8506,
Japan
\and Research Center for Nuclear Physics (RCNP), Osaka University, 
Ibaraki 567-0047, Japan,
\and Department of Physics, Tokyo Metropolitan University, Hachioji 192-0397, Japan}
%
\date{Received: date / Revised version: date}
%
\abstract{
The $\eta$ mesic nucleus is considered to be one of the interesting exotic many body systems 
and has been studied since 1980's theoretically and experimentally.  
Recently, the formation of the $\eta$ mesic nucleus in the fusion reactions
of the light nuclei such as $d + d \rightarrow (\eta + \alpha) \rightarrow X$
has been proposed and the experiments have been performed by
WASA-at-COSY.
We develop a theoretical model to 
evaluate the formation rate of the $\eta$ mesic nucleus in the fusion reactions and show the 
calculated results.  We find that the $\eta$ bound states could be observed in the reactions in cases with the strong attractive and small absorptive $\eta$-nucleus interactions.   
We compare our results with existing data of the $d + d
\rightarrow \eta + \alpha$ and the 
$d + d \rightarrow {^3 \rm He} + N + \pi$ reactions.
We find that the analyses by our theoretical model with the
existing data can provide new information on the $\eta$-nucleus interaction.
} 
\maketitle
\section{Introduction}
\label{intro}
The existence of the bound states of the $\eta$ meson in nucleus 
($\eta$ mesic nuclei) were predicted first by Haider and Liu~\cite{Liu}
in 1980's.
Stimulated by this theoretical result, there have been many studies of
the structure and the formation reactions of the $\eta$ mesic
nucleus~\cite{Chrien:1988gn,Berger:1988ba,Kohno:1989wn,Kohno:1990xv,Chiang:1990ft,Sokol,Johnson:1993zy,Waas:1997pe,Tsushima_Saito,Hayano:1998sy,Inoue:2002xw,GarciaRecio:2002cu,Jido,Nagahiro:2003iv,Pfeiffer:2003zd,Hanhart:2004qs,Nagahiro:2005gf,Kelkar:2006zs,Jido:2008ng,Song:2008ss,Budzanowski:2008fr,Nagahiro:2008rj,Haider:2015fea}.
Recently, the $\eta$-nucleon and $\eta$-nucleus interactions have
been studied theoretically in the context of the chiral symmetry of the
strong interaction and the
$\eta$ mesic nucleus can be considered as one of the interesting objects
to investigate the aspects of the chiral symmetry at finite
density~\cite{Waas:1997pe,Inoue:2002xw,GarciaRecio:2002cu,Jido,Nagahiro:2003iv,Nagahiro:2005gf,Jido:2008ng,Nagahiro:2008rj}.
As for the experimental studies, the first attempt to observe the $\eta$
mesic nucleus was performed by the ($\pi^+, p$) reaction with finite
momentum transfer~\cite{Chrien:1988gn}, 
and the interpretation of the data is still controversial~\cite{Nagahiro:2008rj}.
After that, there were many experimental searches of the bound states as
reported in
Refs.~\cite{Berger:1988ba,Sokol,Johnson:1993zy,Pfeiffer:2003zd,Budzanowski:2008fr}
for example.
The systems with $\eta$ in the light nucleus such as $\eta$-${\rm ^3He}$
state also have been studied
seriously~\cite{Khemchandani:2001kp,Khemchandani:2003dk,Khemchandani:2007ta}
 and the data of the $p+d \rightarrow \eta + {\rm ^3He}$ reaction were studied 
to deduce $\eta - {\rm ^3He}$ interaction~\cite{Xie:2016zhs}. 
So far, the existence of the $\eta$-Mg bound state was concluded in Ref.~\cite{Budzanowski:2008fr}.
However, we have not found any decisive evidence of the 
existence of the $\eta$ bound state in lighter nuclei like He.

Recently, the
new experiments of the $d + d \rightarrow (\eta + \alpha) \rightarrow X$
reaction have been proposed and performed at WASA-at-COSY~\cite{Krzemien:2015fsa,Krzemien:2014qma,Krzemien:2014ywa,Adlarson:2013xg,Skurzok:2016fuv,Skurzok:2011aa,Adlarson:2016dme}.
In the experiments, the formation cross section of the $\eta$ mesic
nucleus in the $\alpha$ particle in the $d + d $ fusion reaction is
planned to be measured by observing the emitted particles 
from the decay of the $\eta$ mesic nucleus 
below the $\eta$ production threshold.
%
%
The formation rate of the emitting particles  
is expected to be enhanced at the resonance energy
of the $\eta$ bound state formation.   
The shape of the observed spectra in Ref.~\cite{Adlarson:2016dme}
were smooth without any clear peak
structures, and 
the upper limit of the $\eta$-nucleus formation cross section was
reported to be 
around 3--6~nb for $d + d \rightarrow {}^{3}\mathrm{He} + n + \pi^{0}$
reaction~\cite{Adlarson:2016dme}.
To evaluate these upper limits, the Fermi motion of $N^{*}$ in nucleus~\cite{Kelkar:2016uwa}
is also taken int account.
The upper limit for 
$ {}^{3}\mathrm{He} + p + \pi^{-}$ 
final state is two times larger because of isospin~\cite{Adlarson:2016dme}.
%
There were also the data of $\eta$ production reaction $d + d \rightarrow
\eta + \alpha$ above the
threshold~\cite{Frascaria:1994va,Willis:1997ix,Wronska:2005wk}, which
are expected to provide the valuable information on the reaction mechanism.

In this paper, we develop a theoretical model to evaluate the
formation cross section of the $\eta$ bound states in the $d + d$
reaction and show the numerical results.   
This theoretical model can be used to deduce the
information on the $\eta$-nucleus interaction from the experimental spectra.
We explain the details of our theoretical model in section~\ref{form}.  
We show the numerical results and compare them with the existing data
in section~\ref{result} to deduce the information on $\eta$-nucleus
interaction, and summarize this paper in section~\ref{concl}.

\section{Formulation}\label{form}

\begin{figure}
\begin{center}
\resizebox{0.4\hsize}{!}{%
 \includegraphics{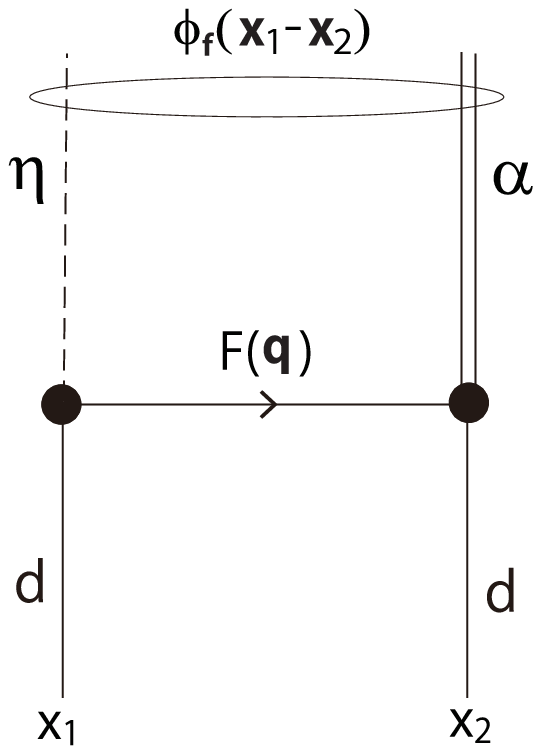}
}
\caption{The schematic diagram of the $d + d \rightarrow \eta + \alpha$
reaction.}
\label{fig:1}       
\end{center}
\end{figure}

In this section, we consider the
$d + d \rightarrow (\eta + \alpha) \rightarrow X$
reaction and explain our theoretical model developed in this article.
In the experiments of this reaction at
WASA-at-COSY~\cite{Krzemien:2015fsa,Krzemien:2014qma,Krzemien:2014ywa,Adlarson:2013xg,Skurzok:2016fuv,Skurzok:2011aa}, 
the total energy of the system is varied by changing the deuteron beam
momentum around the $\eta + \alpha$ threshold energy, which corresponds
to the beam momentum $p=2.3$~GeV/$c$,
and the $\eta + \alpha$ bound state production signal
is expected to be observed as peak structures of the cross section 
of the $d + d \rightarrow {^3 \rm He} + p + \pi^{-}$
and $d + d \rightarrow {^3 \rm He} + n + \pi^{0}$ reactions
in the energy region below the $\eta$-$\alpha$ production threshold
energy region.

Based on these considerations, we have developed a
phenomenological model described below.
In the model, the fusion and $\eta$ meson production processes are
phenomenologically parameterized and the Green's function technique
is used to sum up all $\eta$-$\alpha$ final states.

First we formulate the transition amplitude for the $d+d \to \eta + \alpha$ reaction.
We adopt the framework of the hadron reaction phenomenologically. 
We take a model in which 
the $\eta$ meson production and the $d+d \to \alpha$ fusion take place in a finite size region
as schematically pictured in Fig.~\ref{fig:1}.
All the information on the finiteness of
the reaction range, the spacial dimensions of the nuclear sizes, the structure
of the deuterons and the alpha and overlap of their wavefunctions is 
represented by transition form factor $F(\vec{q})$. 
We are interested in the $\eta$
production at the threshold, so that the final state, $\eta$ and $\alpha$, is dominated 
by $s$ wave and, thus, the total spin-parity of the final state with a
pseudo scalar ($\eta$) and a scalar ($\alpha$) bosons 
is $0^{-}$. Deuteron having spin 1 is represented by a axial vector
boson. 
According to Lorenz invariance,
pseudoscalar $0^-$ state can be made 
out of two axial-vectors $1^{+}$ by so-called anomalous coupling like
$\epsilon^{\mu\nu\rho\sigma} \partial_{\mu} A_{\nu} \partial_{\rho}
A_{\sigma} P S$,
where $A_{\mu}$ and $P$ are an axial-vector boson and a pseudoscalar
boson, respectively. $S$ is a scalar boson.
 Thus, the interaction Hamiltonian 
may be written as
\begin{eqnarray}
  {\cal H}_{\rm int} &=& -i
   c  \epsilon^{ijk}((\partial_{x_{2}^{0}} \nabla^{i}_{x_{1}} -
   \partial_{x_{1}^{0}}\nabla^{i}_{x_{2}})   \hat \phi^{j}_{d}(x_{1})
   \hat \phi^{k}_{d} (x_{2})) \,  \nonumber\\
 &  & \hat \phi_{\eta}^{\dagger}(x_{1}) \hat
 \phi_{\alpha}^{\dagger}(x_{2}) \, {\cal F}(x_{1},x_{2}) 
\label{eq:intH}
\end{eqnarray}
where  
$\hat \phi^{i}_{d}(x)$ is the deuteron field operator with spin index $i$, 
$\hat \phi_{\eta}^{\dagger}(x)$ and $\hat \phi_{\alpha}^{\dagger}(x)$ are 
the creation operators for $\eta$ and $\alpha$, respectively, 
and $c$ expresses the interaction strength.
The interaction strength $c$ will be adjusted so as to reproduce the
observed cross section. 
The function ${\cal F}(x_{1}, x_{2})$ in Eq. (\ref{eq:intH}) represents non-local transition
form factor of $d + d \to \eta + \alpha$, which is supposed to include the information on 
the $d + d \to \alpha$ fusion, and 
the $\eta$ meson production in the hadronic interaction such as $N + N \to \eta + N + N$.
Assuming the translational invariant,
we define the Fourier transformation
\begin{equation}
   {\cal F}(x_{1}, x_{2}) = \int \frac{d^{4}q}{(2\pi)^{4}} F(\vec q) e^{i q\cdot (x_{1} - x_{2})}.
%
\label{eq:FT}
\end{equation}
The momentum transfer $\vec q$ of the reaction is large and 
all nucleons should participate in the fusion reaction equally.
Since it is hard to calculate $F(\vec q)$ in a microscopic way,
we treat it phenomenologically and 
assume a functional form of $F(\vec q)$ in the numerical evaluation.

Letting the wave functions of the incident deuterons labeled by $d_{1}$ and $d_{2}$ 
be given by plane waves with momentum $p_{1}$ for deuteron $d_{1}$ and
$p_{2}$ for deuteron $d_{2}$
and writing the wave functions of $\eta$ and $\alpha$ in the final state as 
$\phi_{\eta}^{\dagger}(\vec x) e^{-iE_{\eta} x^{0}}$ and
 $\phi_{\alpha}^{\dagger}(\vec x) e^{-iE_{\alpha} x^{0}}$
with the $\eta$ and $\alpha$ energies $E_{\eta}$ and $E_{\alpha}$, respectively, 
we obtain the connected part 
of the $S$-matrix in the center of mass frame:
\begin{eqnarray}
   S & = &  -i
     {\cal N}_{d_{1}} {\cal N}_{d_{2}}{\cal N}_{\eta} {\cal N}_{\alpha} 
   c \int d^{4}x_{1} d^{4} x_{2} \epsilon^{ijk} 
  ( \partial_{x_{2}^{0}} \nabla^{i}_{x_{1}} - \partial_{x_{1}^{0}}\nabla^{i}_{x_{2}})
  \nonumber \\ && 
   \left[ \chi_{d_{1}}^{j} \chi_{d_{2}}^{k}
   e^{- i p_{1} \cdot x_{1}} e^{-ip_{2} \cdot x_{2}} 
   + \chi_{d_{2}}^{j} \chi_{d_{1}}^{k}
   e^{- i p_{2} \cdot x_{1}} e^{-ip_{1} \cdot x_{2}} 
   \right]
    \nonumber \\ && \times
   \int \frac{d^{4}q}{(2\pi)^{4}} F(\vec q) e^{i q\cdot (x_{1} - x_{2})}
  \phi^{*}_{\eta}(\vec x_{1}) e^{iE_{\eta}x_{1}^{0}}
   \phi^{*}_{\alpha}(\vec x_{2}) e^{iE_{\alpha} x^{0}_{2}}
\nonumber\\
\label{eq:S}
\end{eqnarray}
where $\chi_{d_{1}}$ and $\chi_{d_{2}}$ are the spin wave functions for 
deuteron $d_{1}$ and $d_{2}$, respectively, 
and the normalization factors ${\cal N}_{i}$ are given as ${\cal N}_{i} = \sqrt{M_{i}/E_{i}}$ 
with mass $M_{i}$ and energy $E_{i}$. 
Operating the derivatives onto the wave functions, we have 
the $S$-matrix as
\begin{eqnarray}
  S &=&  -i 2\pi \delta(E_{1} + E_{2} - E_{f}) 
   {\cal N}_{d_{1}} {\cal N}_{d_{2}}{\cal N}_{\eta}  {\cal N}_{\alpha}  
    \nonumber \\ && \times
      c \int d \vec x_{1} d \vec x_{2} \epsilon^{ijk} 
   2 (E_{2} p_{1}^{i} - E_{1} p_{2}^{i})\chi_{d_{1}}^{j} \chi_{d_{2}}^{k}
   e^{ i \vec p_{1} \cdot \vec x_{1}} e^{i \vec p_{2} \cdot \vec x_{2}} 
    \nonumber \\ && \times
   \int \frac{d \vec q}{(2\pi)^{3}} F(\vec q) e^{-i \vec q\cdot (\vec x_{1} - \vec x_{2})}
  \phi^{*}_{\eta}(\vec x_{1}) \phi^{*}_{\alpha}(\vec x_{2}),
\end{eqnarray}
where $E_{1}$ and $E_{2}$ are the energies for deuteron $d_{1}$ and $d_{2}$, 
respectively, $E_{f} =  E_{\eta} + E_{\alpha}$ and the integrations in terms of 
the time components provide energy conservation
\begin{eqnarray}
&&   \int dx_{1}^{0} dx_{2}^{0} \frac{dq_{0}}{2\pi}
   e^{-ix_{1}^{0}(E_{1}-q^{0} - E_{\eta})} e^{-ix_{2}^{0}(E_{2}+q^{0}-E_{\alpha})}
\nonumber \\ &&
   = 2\pi \delta(E_{1} + E_{2} -E_{f}).
\end{eqnarray}

In order to perform the spacial integrals, we introduce the relative coordinate for 
the final state, $\vec R$ and $\vec r$, defined as
\begin{eqnarray}
   \vec R &=& \frac{m_{\eta} \vec x_{1} + M_{\alpha} \vec x_{2}}{m_{\eta} + M_{\alpha}}, \\
   \vec r & =& \vec x_{1} - \vec x_{2} .
\end{eqnarray}
We also introduce the wave function for the relative motion of the final state, 
$\phi_{f}(\vec r)$, and assume that the center of mass motion of the $\eta$ and $\alpha$ system is written as the plane wave. This implies that we replace the $\eta$ and $\alpha$ wave functions as follows:
\begin{equation}
  {\cal N}_{\eta} {\cal N}_{\alpha} \phi_{\eta}(\vec x_{1}) \phi_{\alpha}(\vec x_{2})
  \to {\cal N}_{f} e^{i \vec p_{f} \cdot \vec R} \phi_{f}(\vec r),
\end{equation}
with the momentum of the center of motion $\vec p_{f} = \vec p_{\eta} + \vec p_{\alpha}$ 
and the normalization of the wave function of relative motion $\phi_{f}(\vec x)$ 
given as
$
   \int d\vec x | \phi_{f}(\vec x)|^{2} = 1.
$
In this coordinate, the $S$-matrix is written as
\begin{eqnarray}
  S &=& -i  2\pi \delta(E_{1} + E_{2} - E_{f}) 
   {\cal N}_{d_{1}} {\cal N}_{d_{2}}{\cal N}_{f}  \nonumber \\ && \times
      c  \epsilon^{ijk} 
   2 (E_{2} p_{1}^{i} - E_{1} p_{2}^{i})\chi_{d_{1}}^{j} \chi_{d_{2}}^{k}
    \nonumber \\ && \times
   \int d \vec R d \vec r\,
   e^{ i (\vec p_{1} + \vec p_{2}) \cdot \vec R} \, e^{i (\frac{M_{\alpha} }{m_{\eta}+M_{\alpha}} \vec p_{1}- \frac{m_{\eta} }{m_{\eta}+M_{\alpha}}\vec p_{2})  \cdot \vec r} 
   \nonumber \\ && \times
   \int \frac{d \vec q}{(2\pi)^{3}} F(\vec q) e^{-i \vec q\cdot \vec r}
  \phi^{*}_{f}(\vec r) e^{-i \vec p_{f} \cdot \vec R} \nonumber  .
\end{eqnarray}
The integral in terms of $\vec R$ provides momentum conservation 
\begin{equation}
   \int d \vec R e^{i(\vec p_{1} + \vec p_{2} - \vec p_{f}) \cdot \vec R} 
   = (2\pi)^{3} \delta (\vec p_{1} + \vec p_{2} - \vec p_{f}). 
\end{equation}
Introducing the Fourier transform
\begin{equation}
   \phi_{f}(\vec r) = \int \frac{d \vec p}{(2\pi)^{3}} \tilde \phi_{f} (\vec p) e^{-i\vec p \cdot \vec r},
\end{equation}
we perform the spacial integrals and obtain
\begin{eqnarray}
   S &=& -i (2\pi)^{4} \delta^{(4)}(p_{1} + p_{2} - p_{f})
      {\cal N}_{d_{1}} {\cal N}_{d_{2}}{\cal N}_{f}  \nonumber \\ && \times
      c  \epsilon^{ijk} 
   2 (E_{2} p_{1}^{i} - E_{1} p_{2}^{i})\chi_{d_{1}}^{j} \chi_{d_{2}}^{k}
    \nonumber \\ && \times
      \int \frac{d \vec q}{(2\pi)^{3}} F(\vec q)  
      \tilde\phi^{*}_{f} (\vec{P}),
\end{eqnarray}
where $\vec{P}$ is defined as $\displaystyle \vec{P} = \frac{M_{\alpha}}{m_{\eta}+M_{\alpha}} \vec p_{1}-
\frac{m_{\eta}}{m_{\eta}+M_{\alpha}} \vec p_{2} - \vec q$.

In the center of mass frame, $E_{1}=E_{2} \equiv E_{d}$ and $\vec p_{1} = - \vec p_{2} \equiv \vec p$.  
Since the $T$ matrix is given by $S = 1 - i T (2\pi)^{4} \delta^{(4)}(p_{1} + p_{2} - p_{f})
{\cal N}_{d_{1}} {\cal N}_{d_{2}} {\cal N}_{f}$, we obtain the $T$-matrix in the center 
of mass fram  as
\begin{equation}
   T =   4 c E_{d}\,  \vec p \cdot ( \vec \chi_{d_{1}} \times  \vec \chi_{d_{2}}) \tilde F(\vec p),
\end{equation}
where we have defined $\tilde F(\vec p)$ as 
\begin{equation}
    \tilde F(\vec p) \equiv  \int \frac{d \vec q}{(2\pi)^{3}} F(\vec q)  
      \tilde\phi^{*}_{f}\left( \vec p - \vec q\right) .  \label{eq:Ftilde}
\end{equation}
With this $T$-matrix the cross section of the fusion with $\eta$ production
can be obtained as
\begin{equation}
  d\sigma = \frac{1}{9} \sum_{\chi_{d_{1}},\chi_{d_{2}}, f}\frac{|T|^{2}}{8 p_{\rm c.m.} E_{d}}
  (2\pi)^{4} \delta^{(4)}(p_{i} - p_{f}) \frac{d \vec p_{f}}{(2\pi)^{3} 2 E_{f}},  
\end{equation}
where $p_{\rm c.m.} = | \vec p|$ and
we take average for the initial spin and sum up all the possible final states.
Performing the integral in terms of $\vec p_{f}$ and taking spin sum, we obtain
\begin{equation}
  \sigma = \frac{2\pi}{9} c^{2} p_{\rm c.m.}
  \sum_{f} |\tilde F(\vec p)|^{2} \delta(E_{i} - E_{f}),
\label{sigma}
\end{equation}
where we have used $E_{i} = 2 E_{d} = E_{f}$.

The total cross section can be written with the Green's function of the
$\eta$ meson. 
Using Eq.(\ref{eq:Ftilde}), we have
\begin{eqnarray}
   \lefteqn{\sum_{f} |\tilde F(\vec p)| ^{2} \delta(E_{f} - E_{i}) } && \nonumber \\
   &=&
   \frac{1}{(2\pi)^{6}} \sum_{f} \delta(E_{f} - E_{i}) 
   \int d \vec q_{1} \, d \vec r_{1} F(\vec q_{1}) e^{-i(\vec q_{1} - \vec p) \cdot \vec r_{1}} \phi^{*}_{f}(\vec r_{1})
   \nonumber \\ && \times
       \int d \vec q_{2} \, d \vec r_{2} F^{*}(\vec q_{2}) e^{i(\vec q_{2} - \vec p) \cdot \vec r_{2}} \phi_{f}(\vec r_{2}).  
\end{eqnarray}
The sum over the final states provides the inclusive spectrum and can be
evaluated by using Green's function method as follows:
Using the formula $\displaystyle \delta(x) = - \frac{1}{\pi} {\rm Im} \frac{1}{x + i\epsilon}$ 
for an infinitesimal quantity $\epsilon$, we obtain
\begin{eqnarray}
 &&  \sum_{f} \delta(E_{f} - E_{i}) \phi_{f}^{*}(\vec r_{1}) \phi_{f}(\vec r_{2})\nonumber \\ 
   &=& - \frac{1}{\pi} {\rm Im} \sum_{f} \phi^{*}_{f}(\vec r_{1}) \frac{1}{E_{f} - E_{i} + i \epsilon} \phi_{f}(\vec r_{2}) \ \ \ \\
   &=& - \frac{1}{\pi} {\rm Im} \sum_{f} \phi^{*}_{f}(\vec r_{1})
    \frac{1}{\hat H - E_{i} + i \epsilon} \phi_{f}(\vec r_{2}),  
\label{eq:Green}
\end{eqnarray}
where we have used the fact that the wave function $\phi_{f}(\vec r)$ is
an eigenfunction of the Hamiltonian $\hat H$ for the $\eta$--$\alpha$ system.  
Equation~(\ref{eq:Green}) is an representation of Green's operator 
$(\hat H - E_{i} + i \epsilon)^{-1}$ 
in terms of the eigenfunction of the Hamiltonian. 
Thus, we write the Green's function in the coordinate space as 
$G(E_{i} ; \vec r_{1}, \vec r_{2})$, and we have 
\begin{eqnarray}
 \sigma &=& \frac{2}{9} c^{2} p_{\rm c.m.} \nonumber\\
 &\times&
 (-) {\rm Im} \int d \vec r_{1} d \vec r_{2}
  f(\vec r_{1}) e^{i \vec p \cdot r_{1}} G(E_{i}; \vec r_{1}, \vec r_{2}) f^{*}(\vec r_{2}) e^{-i \vec p \cdot \vec r_{2}},
\nonumber\\
\end{eqnarray}
where we have introduced the coordinate space expression of $F(\vec q)$ as
\begin{equation}
  f( \vec r) \equiv \int \frac{d\vec q}{(2\pi)^{3}} F(\vec q) e^{- i \vec q \cdot \vec r}.
\label{eq:F_space}
\end{equation}

Further by assuming the spherically symmetric form of $f$ for simplicity,
we replace $f({\vec r})$ as,
\begin{eqnarray}
 f({\vec r}) \rightarrow f(r).
\end{eqnarray}
And by making use of the multipole expansion of $G$,
\begin{eqnarray}
 G(E_{i} ; {\vec r_1}, {\vec r_2})  
= \sum_{\ell m} Y_{\ell m}(\hat{r}_1) Y_{\ell m}^{*}(\hat{r}_2)
G^{\ell} (E_{i} ; r_1, r_2), 
\end{eqnarray}
we obtain the final form of $\sigma$ as,
\begin{eqnarray}
 \sigma &=& - \frac{8\pi}{9} c^{2} p_{\rm c.m.} \ {\rm Im}
 \int r_{1}^2 dr_1 r_{2}^2 dr_2 \ f(r_1) f^{*}(r_2)  \nonumber\\ 
 &\times& \sum_{\ell} (2 \ell +1) \ j_{\ell}(p r_1)
\ G^{\ell} (E_{i} ;  r_1,  r_2) \ j_{\ell}(p r_2),
\label{eq:sigma}
\end{eqnarray}
where $j_{\ell}$ is the spherical Bessel function.
This expression 
is used for the numerical evaluation of the fusion cross section
of Eq.~(\ref{sigma}).

We can divide the total cross section $\sigma$ into two parts, 
the conversion part $\sigma_{\rm conv}$ and the escape part $\sigma_{\rm esc}$ 
as,
\begin{equation}
 \sigma = \sigma_{\rm conv} + \sigma_{\rm esc},
\label{eq:sigma_conv_esc}
\end{equation}
based on the identity
\begin{equation}
 {\rm Im}G = \{G^{\dagger} {\rm Im} U G \}+ \{(1+G^{\dagger}U^{\dagger}) {\rm Im}G_{0} (1+UG) \},
\label{eq:ImG}
\end{equation}
where $G_{0}$ is the free Green's function of $\eta$ and $U$ the
$\eta$-nucleus potential~\cite{Green}.
The first term of the R.H.S of Eq.~(\ref{eq:ImG})
represents the contribution of the $\eta$ meson absorption by the
imaginary part of the $\eta$-nucleus potential $U$ and is called as the
conversion part.
The second term is the contribution from the $\eta$ meson escape from
the nucleus and is called as the escape part.  

The conversion part of the cross section $\sigma_{\rm conv}$ is evaluated
by the practical form written as,
\begin{eqnarray}
\sigma_{\rm conv} &=&
- \frac{8\pi}{9} c^{2} p_{\rm c.m.} 
\int r_{1}^2 dr_1 r_{2}^2 dr_2 r_{3}^2 dr_3
\ f(r_1) f^{*}(r_3) \nonumber\\ 
 &\times&  {\rm Im} U_{\rm opt}(r_{2})
  \sum_{\ell}(2\ell +1) j_{\ell}(p r_1) \nonumber\\ 
&\times&  
 G^{\ell *} (E_{i} ;  r_1,  r_2) \ G^{\ell} (E_{i} ; r_2, r_3)
 \ j_{\ell}(p r_3),
\label{eq:21}
\end{eqnarray}
for the spherical $\eta$-$\alpha$ optical potential $U_{\rm opt}(r)$.
We calculate the escape part as
$\sigma_{\rm esc} = \sigma - \sigma_{\rm conv}$.

The conversion and escape parts have the different energy dependence.
In the subthreshold energy region of the $\eta$ meson production,
the total cross section $\sigma$ is equal to the conversion part
$\sigma_{\rm conv}$ since the energy of the $\eta$ meson is insufficient
to escape from the nucleus and all $\eta$ mesons must be absorbed to the
nucleus finally.
Thus, the signal of the formation of the $\eta$ bound state is expected
to be observed in $\sigma_{\rm conv}$.
As shown in Eq.~(\ref{eq:21}), the expression of the conversion cross
section $\sigma_{\rm conv}$ includes the Green's function $G^{\ell}$
which is responsible for the peak and cusp structures in the spectrum 
as the consequences of the $\eta$-nucleus interaction such as the bound
state formation with angular momentum $\ell$.
For higher partial waves $\ell$ without any bound states, 
the energy dependence of $G^{\ell}$ is tend to be rather mild and almost
flat in the energy region of the $\eta$ production threshold.
In addition, $\sigma_{\rm conv}$ also includes ${\rm Im} U_{\rm opt}$
and, thus, the size of $\sigma_{\rm conv}$ will be larger for stronger
absorptive potential.
Consequently, as shown later, the flat contribution to the
conversion spectrum gets larger in proportional to the
strength of the absorptive potential.
Above the threshold of $\eta$ production,
we have also the contribution from the escape processes.
This escape part $\sigma_{\rm esc}$ can be compared to the observed $\eta$
production cross section in the $d + d \rightarrow \eta + \alpha$
reaction.

We should mention here to the effects of the distortion of the initial
deuterons to the calculated results.
In Eq.~(\ref{eq:S}), the deuteron waves are introduced to the formula as
plane waves. The distortion effects will modify the deuteron-deuteron
relative wave function and could change the results. In the present
cases, however, we can expect the effects will be minor in the final
spectra shown in next section by the following reasons.
The energy range of the final spectra 
considered in this article is very narrow and restricted to only around
the $\eta$ production threshold. Thus, the distortion effects between 
two deuterons are almost constant
in this narrow energy range and are expected to change only absolute
value of the spectra by an almost constant factor.
On the other hand, in the present analyses, we normalize the calculated
results using the experimental data of $d+ d \rightarrow \eta + \alpha$ 
reaction observed above the threshold as we will see later.
Thus, the final results are expected to be insensitive to the deuteron
distortion.
We have checked qualitatively this statement
by introducing the spherical distortion factor to suppress the
contributions from the small relative coordinate region
in Eqs.~(\ref{eq:sigma}) and (\ref{eq:21}),
and we have confirmed that the distortion effects to the spectra is
almost constant within the accuracy of around 15~\% in the energy region
considered here.
Hence, we can neglect the deuteron distortion effects in this
article.

As for the numerical evaluation,  we assume the $\eta$-$\alpha$ optical potential
has the following form,
\begin{eqnarray}
 U_{\rm opt} (r) = (V_0 + i W_0)\frac{\rho_{\alpha}(r)}{\rho_{\alpha}(0)},
\label{eq:Uopt}
\end{eqnarray}
where $V_0$ and $W_0$ are the parameters to determine the potential
strength at center of the $\alpha$ particle.
The density of the $\alpha$ particle
$\rho_\alpha (r)$ is assumed to have Gaussian form, 
\begin{equation}
 \rho_{\alpha}(r) = \rho_{\alpha}(0) \exp\left[ - \frac{r^2}{a^2}\right],
\label{eq:rho}
\end{equation}
with the range parameter $a=1.373$~fm
which reproduces the R.M.S radius of $\alpha$ to be 1.681 fm. 
The central density of this distribution is $\rho_{\alpha}(0) =
0.28~{\rm fm}^{-3} \simeq 1.6 \rho_0$ with the normal nuclear density
$\rho_0=0.17$~fm$^{-3}$.
As the practical form of $F$, we assume the Gaussian as,
\begin{eqnarray}
 F({\vec p}) = (2\pi)^{3/2} \left(\frac{2}{p_{0}^2 \pi} \right)^{3/4} 
 \exp\left[ - \frac{p^2}{p_{0}^2}\right],
\label{eq:F}
\end{eqnarray}
in this article and we treat $p_0$ as a phenomenological parameter.
We also show the numerical results obtained by choosing 
other functional forms for the transition form factor $F({\vec p})$
($f({\vec r})$) in Appendix to estimate the functional form dependence
of our results.

We make a few comments on the 
difficulties for developing more microscopic model to evaluate the reaction rate. 
The first difficulty is the large momentum transfer.
We need around 1~GeV/$c$ momentum transfer 
at the $\eta$ production threshold in the center of mass frame.
The accuracy of the microscopic wave function in such high
momentum transfer region is not well investigated.
Another difficulty arises from the fact that the reaction is a fusion
reaction. In the fusion reaction, all particles in the system participate 
the reaction equally and receive large momentum transfer.
Because of these features of this reaction, we need the sufficiently 
accurate wave function of the five-body system (four nucleons and one
$\eta$ meson) 
and the reliable description of the fusion and $\eta$ production processes  
to perform the fully microscopic calculation.

On the other hand, at the same time,
we can also find an advantage for the studies of
this reaction.
We can make use of the experimental data of the
$d +d \rightarrow \eta + \alpha$ reaction just above the
$\eta$ production
threshold~\cite{Frascaria:1994va,Willis:1997ix,Wronska:2005wk}.
These data must provide us important information on the reaction 
and can be
used to fix the parameters included in the present model.
It should be also mentioned that the energy spectrum of the $\eta$-$\alpha$
system is expected to be simple since the system is small
and may have only a few bound levels of $\eta$ even if they exist.
The simple spectrum could be helpful to identify the bound levels from the data.

\section{Numerical Results and Discussions}\label{result}

\begin{figure}
\begin{center}
\resizebox{0.95\hsize}{!}{%
 \includegraphics{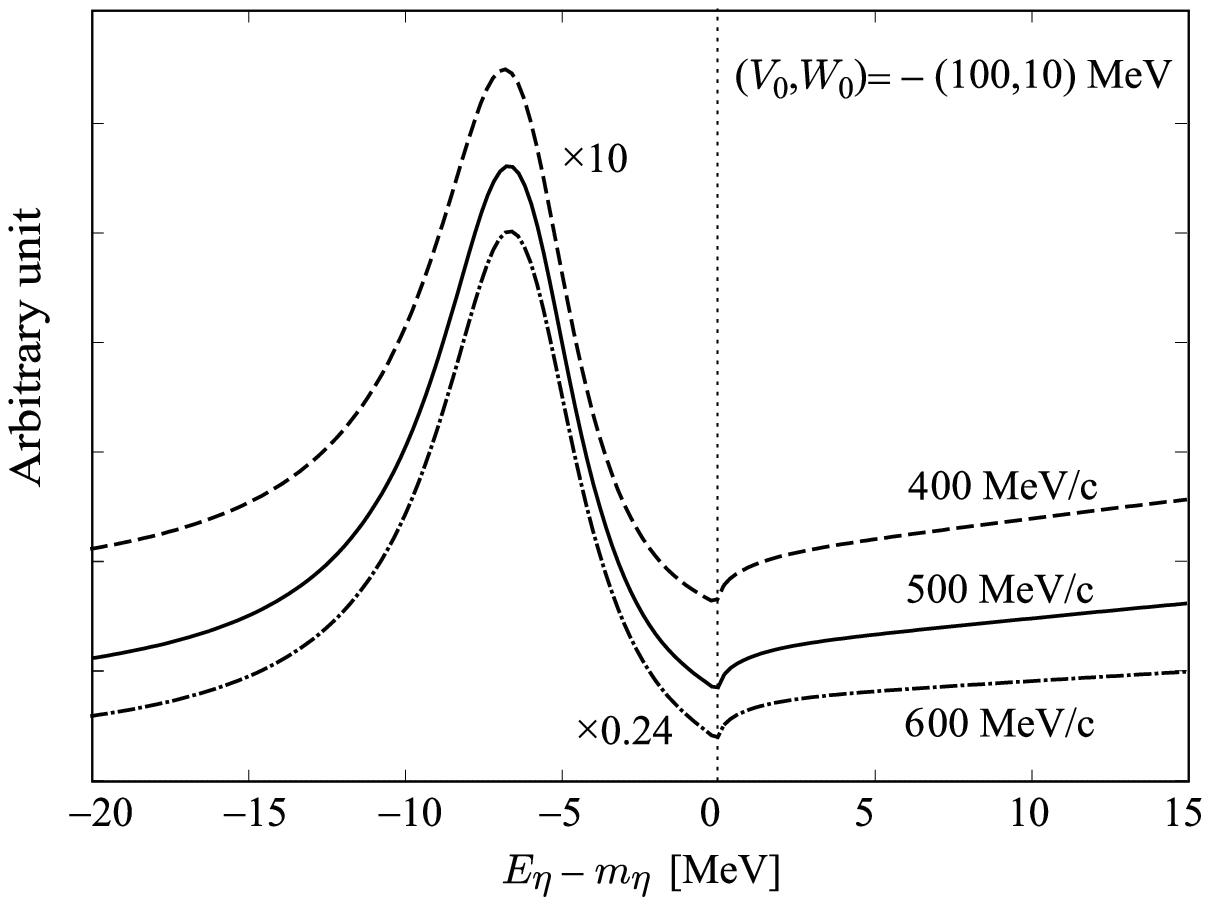}}
\caption{Calculated total cross sections 
of the $d +d \rightarrow ( \eta + \alpha ) \rightarrow X$ reaction for
 the formation of the $\eta$-$\alpha$ bound system 
with $p_0 = 400, 500, 600$ MeV/$c$ cases
plotted as functions of the $\eta$ excited energy $E_\eta - m_\eta$.
The parameters of the $\eta$-$\alpha$ optical potential are fixed to be
$(V_0, W_0) = -(100, 10)$ MeV.
It is noted that the result with $p_0 =600$ MeV/$c$ is scaled by
 factor $0.24$ and that with $p_0 = 400$ MeV/$c$ scaled by factor 10.
}
\label{fig:2.1}       
\end{center}
\end{figure}

The theoretical model described in section~\ref{form} includes three
parameters, which are the strength of the real and imaginary parts of
the $\eta$-$\alpha$ potential ($V_0, W_0$)
defined in Eq.~(\ref{eq:Uopt}), and the parameter $p_0$ appeared in 
Eq.(\ref{eq:F}) to determine the property of the function $F$ which
physical meanings are explained in Eqs.~(\ref{eq:intH}) and (\ref{eq:FT}).

We study first the sensitivity of the shape of the cross section to the
parameter $p_0$.
We show the calculated total cross section $\sigma$ in Eq.~(\ref{eq:sigma})
for $p_0 = 400, 500, 600$ MeV/$c$ cases with $(V_0, W_0) = -(100, 10)$ MeV 
as functions of the $\eta$ excited energy $E_\eta - m_\eta$ 
in Fig.~\ref{fig:2.1}.
The peak structure of the results with $p_0 = 400$ MeV/$c$ is small 
and is located on the top of almost flat spectrum. 
We find that 
the structures appearing in the cross
section is insensitive to $p_0$ and almost same for three $p_0$ values.
Thus, we fix the value of this parameter to be $p_0 = 500$ MeV/$c$ in the 
following numerical results
and focus on the sensitivity of the structure of the spectrum
to the $\eta$-$\alpha$ potential strength.

The parameter $p_0$ dependence of the total cross section can be
understood by considering
the change of spatial dimension of the $f(\vec{r})$ defined in
Eq.~(\ref{eq:F_space}).
For smaller $p_0$ value, the distribution of $F$ in the momentum space
is more compact and that of $f$ in coordinate space is wider.
Thus, for smaller $p_0$ values, we have relatively larger contributions of
higher partial wave $\ell$ of the $\eta$ meson in
the calculation of the cross section in Eq.~(\ref{eq:sigma}), which 
are expected not to 
have any structures as function of energy around the threshold
as mentioned before.
Hence, for smaller $p_0$ values,
the small peak structure appears on the top of the almost flat
contribution in the spectrum.

\begin{figure}
\begin{center}
\resizebox{0.85\hsize}{!}{%
\includegraphics{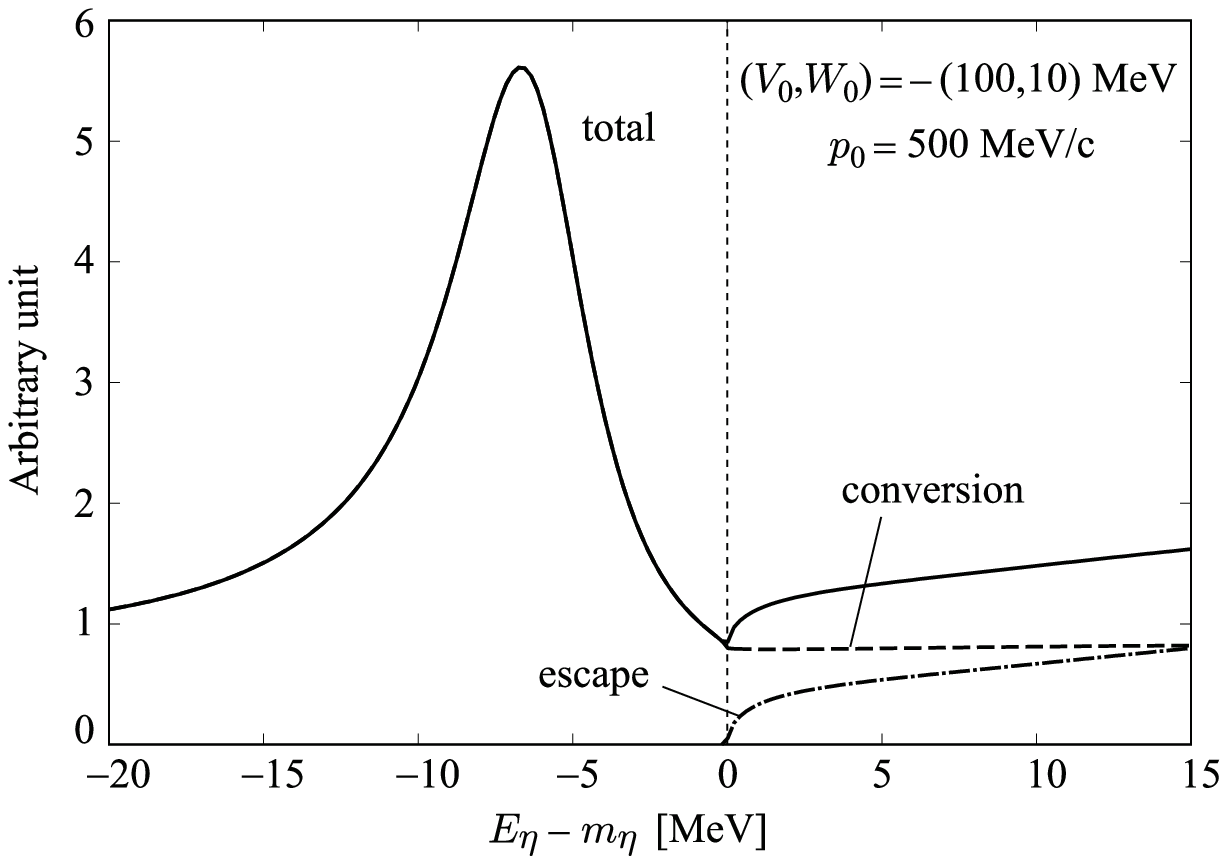}}
\caption{Calculated cross section of the $d + d \rightarrow (\eta +
 \alpha) \rightarrow X$ reaction for the formation of the $\eta - \alpha$ bound
 system plotted as
 functions of the $\eta$ excited energy $E_\eta - m_\eta$. The parameters of the
 $\eta$-$\alpha$ optical potential are $(V_0, W_0) = -(100,10)$ MeV, and
 the parameter $p_0$ is fixed to be $p_0 = 500$ MeV/$c$. 
The thick solid line indicates the total cross
 section $\sigma$. The dashed line and the dot-dashed line indicate the
 conversion part $\sigma_{\rm conv}$ and the escape part $\sigma_{\rm
 esc}$, respectively.
}
\label{fig:2}
\end{center}
\end{figure}

\begin{figure}
\begin{center}
\resizebox{0.85\hsize}{!}{%
 \includegraphics{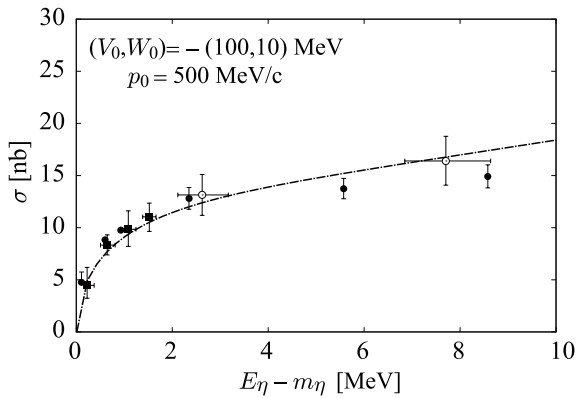}}
\caption{Calculated escape part $\sigma_{\rm esc}$ in Fig.~\ref{fig:2} plotted with the
 experimental data of $d + d \rightarrow
 \eta + \alpha$ reaction indicated by black
 squares~\cite{Frascaria:1994va}, black circles~\cite{Willis:1997ix}, and open
 circles~\cite{Wronska:2005wk}. 
The parameters of the $\eta$-$\alpha$ optical potential are $(V_0, W_0) =
 -(100, 10)$ MeV and the parameter $p_0$ is fixed to be $p_0 =500$
 MeV/$c$. 
The height of the calculated spectrum is adjusted so as to reproduce the
 data by changing the interaction strength $c$ given in Eq.~(\ref{eq:intH}).
}
\label{fig:3}       
\end{center}
\end{figure}

In Fig.~\ref{fig:2}, we show again the calculated $\sigma$ 
for the case with parameters $(V_0, W_0)=-(100,10)$ MeV
for $p_0 = 500$ MeV/$c$ to study the detail structure of the spectrum.
Three lines correspond to the total cross section $\sigma$, 
the conversion part $\sigma_{\rm conv}$,
and the escape part $\sigma_{\rm esc}$.
The $\eta$ production threshold corresponds to $E_\eta - m_\eta =0 $ and
the $\eta$-$\alpha$ bound states are expected to be produced in the
subthreshold region $E_\eta - m_\eta < 0$.
We can see in Fig.~\ref{fig:2} that 
the spectrum has non-trivial structures above the flat contribution whose
height seems to be about 1 in the scale of the vertical axis. 
There is a clear peak at 
$E_\eta - m_\eta \simeq -7$ MeV which corresponds to the formation of
the $\eta$-$\alpha$ bound state.
The calculated escape part
is plotted with data in Fig.~\ref{fig:3}.
We have adjusted the height of the spectrum by the interaction strength $c$
in Eq.~(\ref{eq:intH}).
The agreement of the spectrum shapes of the
calculated results and the data above the
threshold seems reasonably good in this potential parameter.
Hence, this observation implies that
this parameter set, which predicts the formation of
$\eta$-$\alpha$ bound state, does not contradicts to the $d + d \rightarrow
\eta + \alpha$ data above threshold.
We will make further comments for the comparison with the subthreshold data later
in this section.

\begin{figure}
\begin{center}
\resizebox{0.95\hsize}{!}{%
 \includegraphics{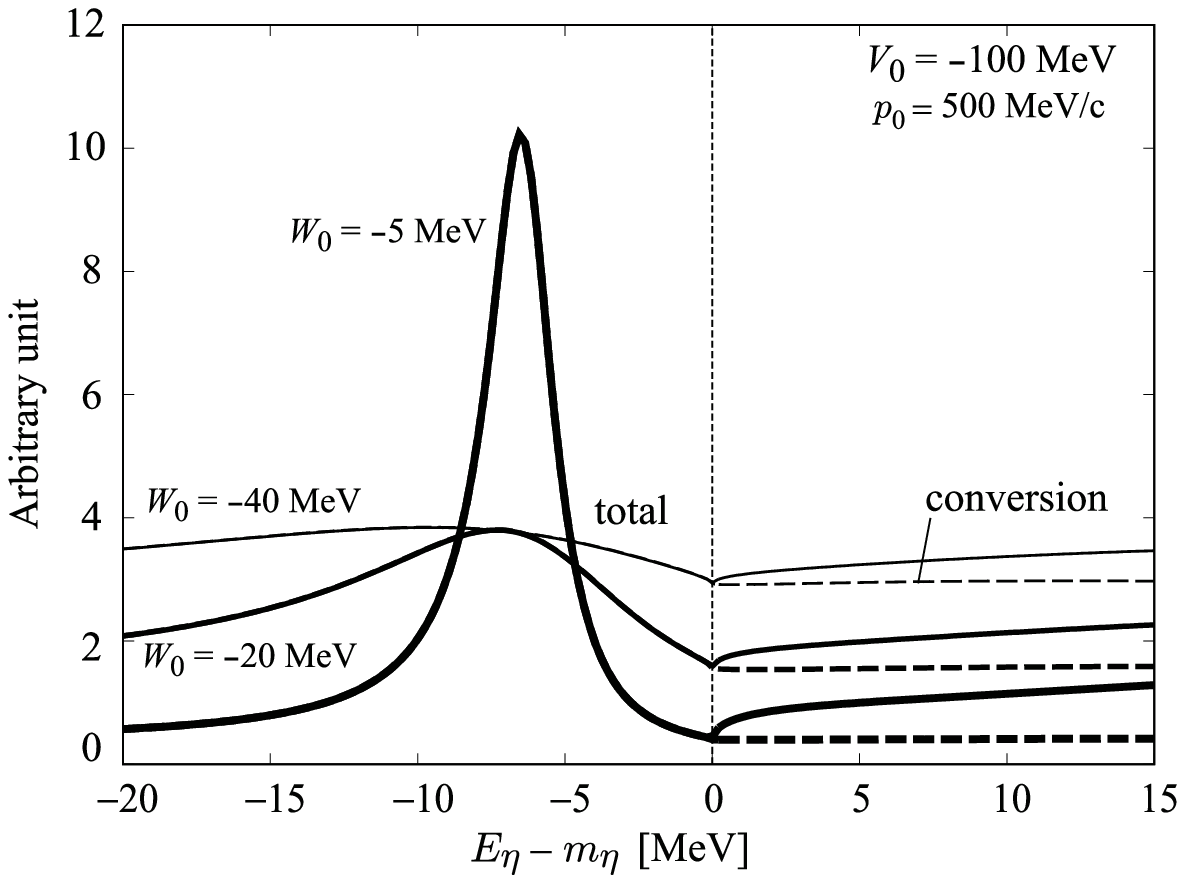}
}
\caption{Calculated cross sections of the $d + d \rightarrow (\eta +
 \alpha) \rightarrow X$ reaction for the formation of the $\eta - \alpha$ bound
 system plotted as
 functions of the $\eta$ excited energy $E_\eta - m_\eta$. The parameters of the
 $\eta$-$\alpha$ optical potential are $(V_0, W_0) = -(100,5), -(100,20)$,
 and $-(100,40)$ MeV, and
 the parameter $p_0$ is fixed to be $p_0 = 500$ MeV/$c$. 
The solid lines indicate the total cross
 sections $\sigma$ and the dashed lines the conversion parts
 $\sigma_{\rm conv}$.}
\label{fig:total_V100_W_p500}       
\end{center}
\end{figure}

\begin{figure}
\begin{center}
\resizebox{0.95\hsize}{!}{%
 \includegraphics{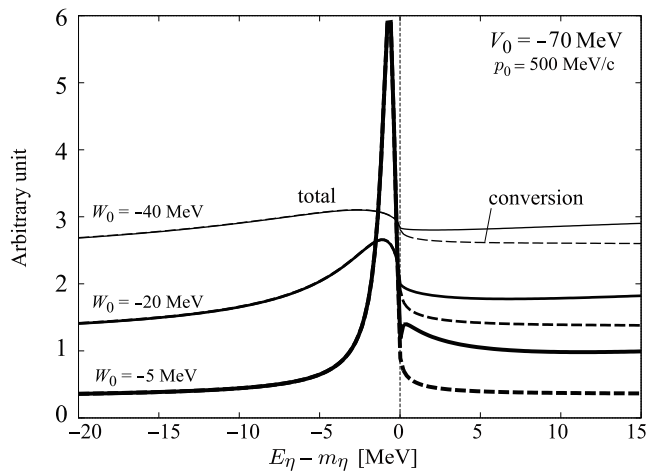}}
\caption{Same as Fig.~\ref{fig:total_V100_W_p500} except for $(V_0, W_0) =
 -(70,5), -(70,20)$, and $-(70, 40)$ MeV.}
\label{fig:total_V70_W_p500}       
\end{center}
\end{figure}

\begin{figure}
\begin{center}
\resizebox{0.95\hsize}{!}{%
 \includegraphics{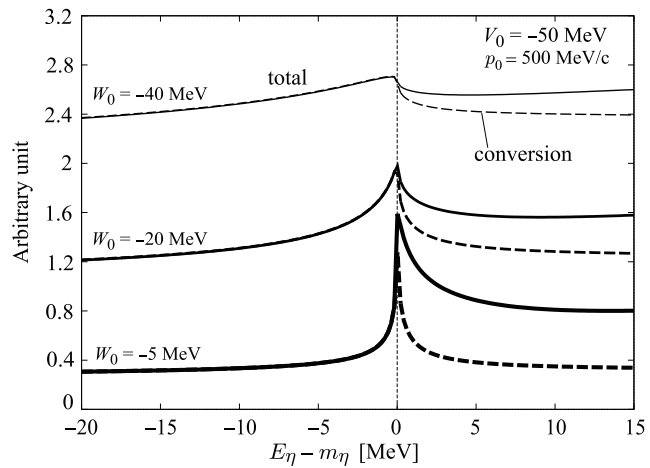}}
\caption{Same as Fig.~\ref{fig:total_V100_W_p500} except for $(V_0, W_0) =
 -(50,5), -(50,20)$, and $-(50, 40)$ MeV.}
\label{fig:total_V50_W_p500}       
\end{center}
\end{figure}

\begin{figure}
\begin{center}
\resizebox{0.95\hsize}{!}{%
  \includegraphics{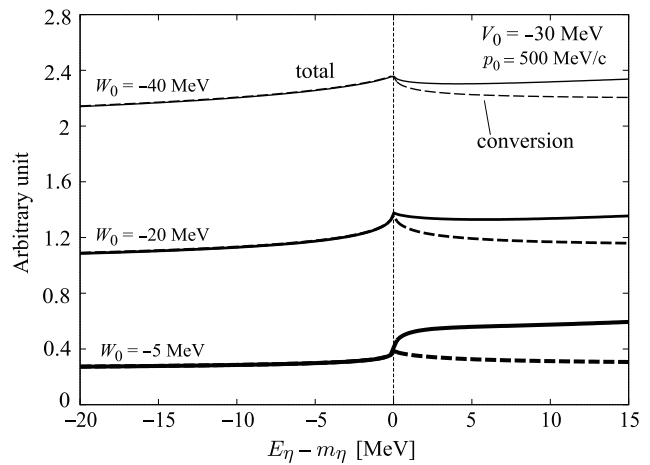} }
\caption{Same as Fig.~\ref{fig:total_V100_W_p500} except for $(V_0, W_0) =
 -(30,5), -(30,20)$, and $-(30, 40)$ MeV.}
\label{fig:total_V30_W_p500}       
\end{center}
\end{figure}

%

In Fig.~\ref{fig:total_V100_W_p500}, 
we show the results for $(V_0, W_0)=-(100,5)$, $-(100,20)$ and 
$-(100,40)$~MeV cases with $p_0 = 500$~MeV/$c$ to see the effects of the
strength of the imaginary potential $W_0$ to the spectra.
We can see from the figures 
the structure of the spectra below threshold $E_\eta -m_\eta < 0$ is
sensitive to the value of the potential parameters $W_0$ and
is expected to be the good observables to investigate the $\eta$-nucleus
interaction.
Actually, the subthreshold spectrum with small imaginary potential $W_0 =
-5$ MeV clearly shows the existence of the bound state around $E_\eta -
m_\eta = -7$ MeV as a peak structure.
The width of the peak becomes wider and the peak height lower for larger
$|W_0|$ value.
At the same time we can see that 
the structure of the spectra above the threshold $E_\eta - m_\eta > 0$
are relatively insensitive to the 
imaginary part of the $\eta$-nucleus interaction and the
value of $W_0$. 

We also show the calculated results for different $V_0$ and $W_0$ values
in Figs.~\ref{fig:total_V70_W_p500},~\ref{fig:total_V50_W_p500} and
\ref{fig:total_V30_W_p500}.
It could be interesting to note that the depth of the $\eta$-nucleus potential by the
chiral unitary model in Refs.~\cite{Inoue:2002xw,GarciaRecio:2002cu}
is roughly close to $-(50,40)$ MeV at normal nuclear density
for real and imaginary parts, respectively.
The depth of the so-called $t\rho$ potential evaluated by using the 
$\eta$-$N$ scattering length $a_{\eta N} = (0.28 + 0.19 i)$ fm
in Ref.~\cite{Bhalerao:1985cr} is around $ - (30,20)$ MeV at normal
nuclear density.
It should be noted that 
the parameters $V_0$ and $W_0$ adopted here in Eq.~(\ref{eq:Uopt}) indicate the
potential strength at the center of the $\alpha-$particle where 
$\rho_\alpha (0) \simeq 0.28$~fm$^{-3}$ as defined in Eq.~(\ref{eq:rho}).
The $\eta$-$\alpha$ potential also has been studied microscopically in 
Refs.~\cite{Rakityansky:1996gw,Kelkar:2007pn,Fix:2017ani}, 
where the various values of the $\eta$-$\alpha$ scattering length
$a_{\eta \alpha}$ were reported.
We could compare our potential strength ($V_0, W_0$) with the scattering
length simply by the Born approximation as
$ \displaystyle a_{\eta \alpha} = -2 \mu \int dr r^2 U_{\rm opt}(r)$
with the $\eta$-$\alpha$ reduced mass $\mu$.
For example, based on this relation,
some of the potential parameters used in this
article correspond to the scattering length as,
\begin{description}
\item $(V_0, W_0)=-(100,5)$~MeV \ $\leftrightarrow$ \ $ a_{\eta \alpha} = 2.81 + 0.14 i$~fm
\item $(V_0, W_0)=-(70,20)$~MeV \ $\leftrightarrow$ \ $ a_{\eta \alpha} = 1.97 + 0.56 i$~fm
\item $(V_0, W_0)=-(50,40)$~MeV \ $\leftrightarrow$ \ $ a_{\eta \alpha} = 1.40 + 1.12 i$~fm.
\end{description}
It will be interesting to compare these numbers with the microscopic
scattering lengths, for example, results of the microscopic calculation
listed in Table~IV in Ref.~\cite{Fix:2017ani}.
We can see that the potential strengths adopted in this article are
within the range of uncertainties of microscopic calculation.

We find again the same tendencies in these results as in Fig.~\ref{fig:total_V100_W_p500}.
In Fig.~\ref{fig:total_V70_W_p500} for $V_0 = -70$~MeV case, we also find the bound state peak at
$E_\eta - m_{\eta} \simeq -1 $~MeV which is very clear for small $|W_0|$
case.
In the result shown in Fig.~\ref{fig:total_V50_W_p500} for $V_0 = -50$
MeV case, we find the cusp structure at the threshold energy, which
becomes less prominent for larger absorption potential.
In Fig.~\ref{fig:total_V30_W_p500}, 
for weaker attractive cases with $V_0 = -30$ MeV,
we find the step like structure at the threshold for weak absorptive
case with $W_0 =-5$ MeV, which becomes less prominent again for stronger
absorptive potential with larger $|W_0|$ value.
From these figures,
we find that the total spectra around and below threshold 
$E_\eta - m_\eta \leq 0$ are 
sensitive to both the real and imaginary parts of the $\eta$-nucleus interaction described by
the parameters ($V_0, W_0$) and are good observables to obtain information on
the $\eta$-nucleus interaction.

As shown in the conversion spectra in
Figs.~\ref{fig:total_V100_W_p500}--\ref{fig:total_V30_W_p500},
we have the almost flat contributions in whole energy region as
mentioned in Sect.~\ref{form}.
This flat contribution is considered to be a part of the background
cross sections of the experimental data.
Thus, we subtract the flat contribution in the conversion part in the
following numerical results to investigate the structure appearing in the
spectrum.
For this purpose, we subtract the minimum value of the conversion cross
section in the energy range shown in the following figures.

\begin{figure}
\begin{center}
\resizebox{0.95\hsize}{!}{%
 \includegraphics{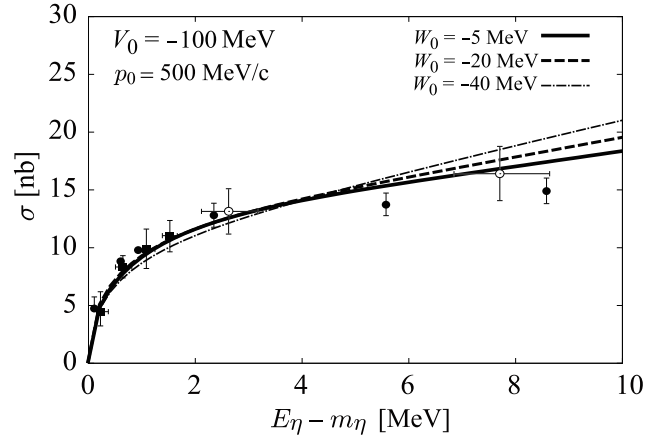} }
\caption{ 
Calculated escape part $\sigma_{\rm esc}$ plotted with the experimental
 data of $d + d \rightarrow \eta + \alpha$ reaction indicated by black
 squares~\cite{Frascaria:1994va}, black circles~\cite{Willis:1997ix}, and open
 circles~\cite{Wronska:2005wk}. 
The parameters of the $\eta$-$\alpha$ optical potential are 
$(V_0, W_0) =  -(100,5), -(100,20)$,
 and $-(100,40)$ MeV as shown in the figure.
The parameter $p_0$ is fixed to be $p_0 =500$~MeV/$c$. 
}
\label{fig:V100_W_p500_w_data}       
\end{center}
\end{figure}

\begin{figure}
\begin{center}
\resizebox{0.95\hsize}{!}{%
 \includegraphics{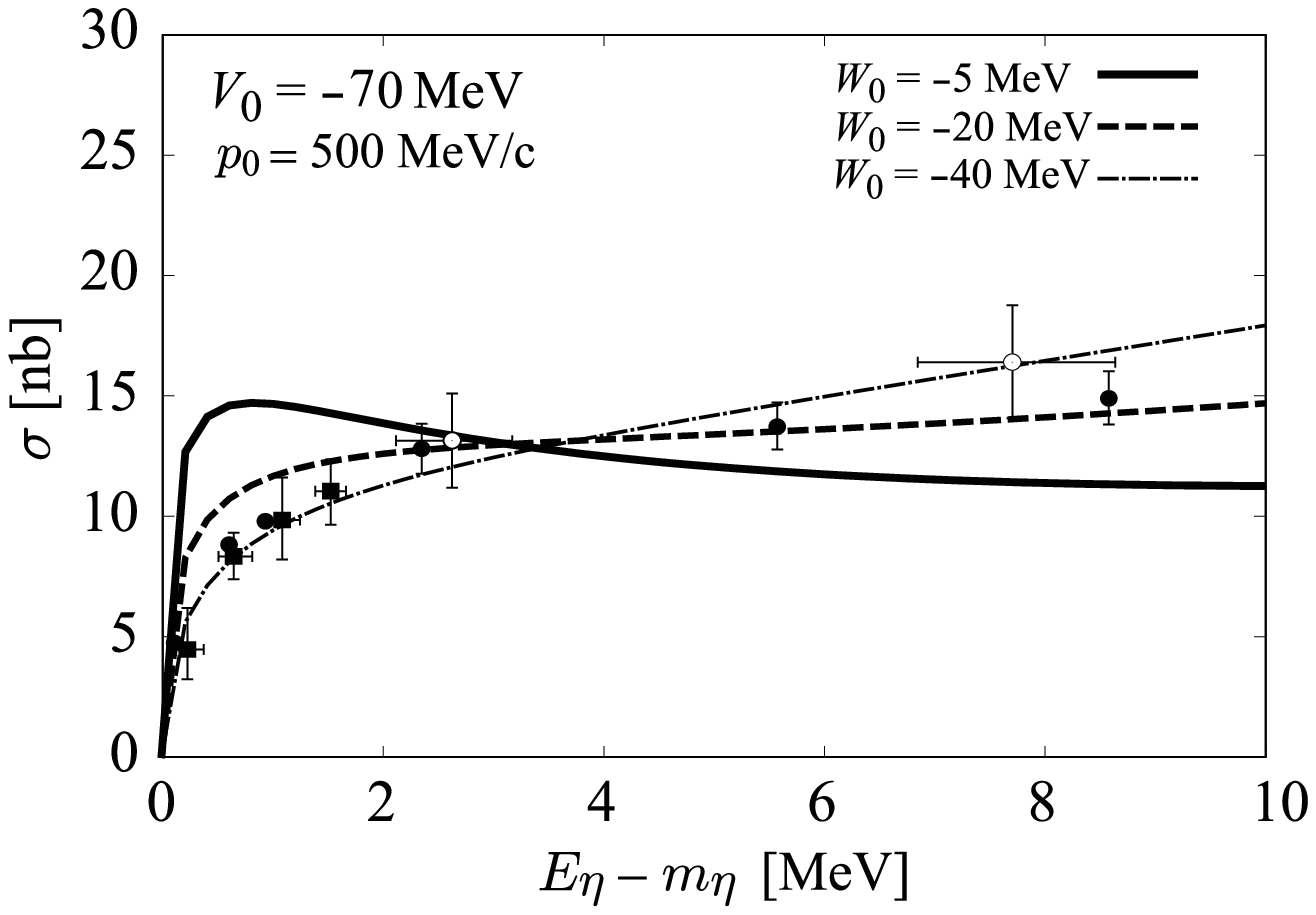} }
\caption{
Same as Fig.~\ref{fig:V100_W_p500_w_data} except for the $\eta$-$\alpha$ optical potential
 parameters are $(V_0, W_0) = -(70,5), -(70,20)$,
 and $-(70,40)$ MeV as shown in the figure. }
\label{fig:V70_W_p500_w_data}       
\end{center}
\end{figure}

\begin{figure}
\begin{center}
\resizebox{0.95\hsize}{!}{%
 \includegraphics{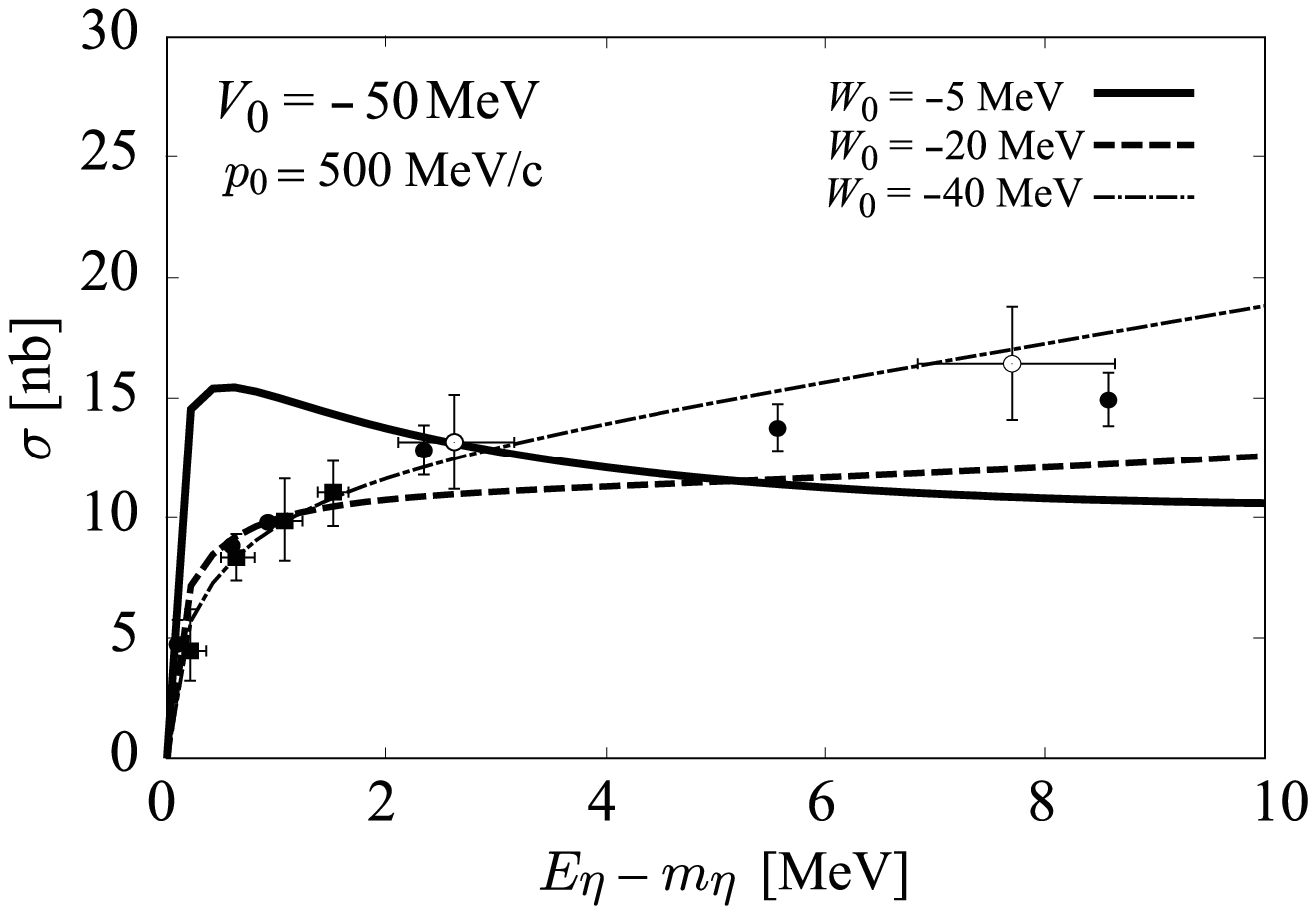} }
\caption{
Same as Fig.~\ref{fig:V100_W_p500_w_data} except for the $\eta$-$\alpha$ optical potential
 parameters are $(V_0, W_0) = -(50,5), -(50,20)$,
 and $-(50,40)$ MeV as shown in the figure. }
\label{fig:V50_W_p500_w_data}       
\end{center}
\end{figure}

\begin{figure}
\begin{center}
\resizebox{0.95\hsize}{!}{%
  \includegraphics{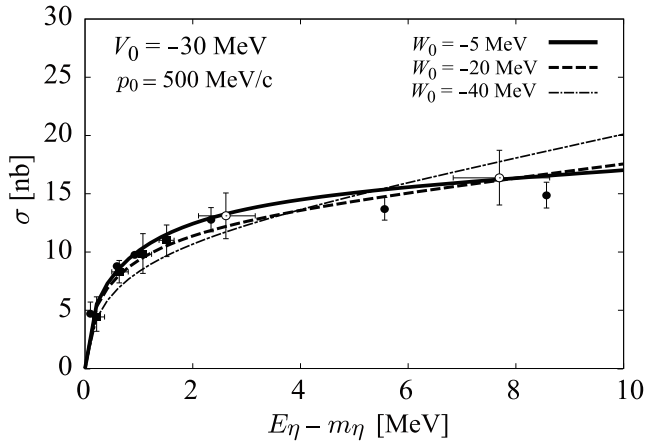}}
\caption{
Same as Fig.~\ref{fig:V100_W_p500_w_data} except for the $\eta$-$\alpha$ optical potential
 parameters are $(V_0, W_0) = -(30,5), -(30,20)$,
 and $-(30,40)$ MeV as shown in the figure.}
\label{fig:V30_W_p500_w_data}       
\end{center}
\end{figure}

It will be extremely interesting to show the binding energies and widths of
the $\eta$-$\alpha$ bound states obtained by solving the Klein-Gordon
equation using the same $\eta$-$\alpha$ potential used to calculate the
spectra in Figs.~\ref{fig:total_V100_W_p500}--\ref{fig:total_V30_W_p500}.
The results are compiled in Table~\ref{BE}.
We find that only peak structures appearing in the spectra for the
strong attractive--weak absorptive potential cases correspond to the
existence of the bound states.
Other structures in the spectra may not indicate the existence of the
bound state, though they provide important information on the
$\eta$-$\alpha$ interaction definitely.

\begin{table}[htb]
\caption{Calculated binding energies (B.E.)
and widths ($\Gamma$)
of the $\eta$-$\alpha$ bound states by solving the Klein-Gordon equation
 with the optical potential defined in Eq.~(\ref{eq:Uopt}).
All states shown here are ground states.
We have not found any bound states for $V_0 = -50$ and $-30$~MeV with
$W_0= -5, -20, -40$~MeV cases.
\label{BE}}
\begin{center}
\begin{tabular}{c|cc|cc} \hline\hline
\multicolumn{1}{c|}{$W_0$}%
&\multicolumn{2}{c|} {$V_0 = -100$~MeV} 
&\multicolumn{2}{c} {$V_0 = -70$~MeV} \\ 
\multicolumn{1}{c|}{[MeV]} & B.E. & ${\rm \Gamma}$  &
 B.E.  & ${\rm \Gamma}$   \\ \hline
$-5$ & 6.4 & 2.7  & 0.5 & 1.1  \\
$-20$& 5.7 & 11.0 &  -   &   -   \\ 
$-40$& 3.3 & 22.9 &  -   &   -   \\ \hline\hline	   
\end{tabular}
\end{center}
\end{table}

Now, we focus on the escape part of the spectrum which appears
above the threshold, $E_\eta - m_\eta > 0$, and can be compared to the experimental
data of the $d + d \rightarrow \eta + \alpha$ reaction as already shown 
in Fig.~\ref{fig:3}
for a certain parameter set.
In Fig.~\ref{fig:V100_W_p500_w_data}, we show the calculated escape part
of the spectra for $V_0 = -100$ MeV cases with different
strengths of the absorptive potential together with the experimental data.
The calculated results are scaled to fit the experimental data
by changing the interaction strength $c$.
As we can see from Figs.~\ref{fig:3} and \ref{fig:V100_W_p500_w_data},
the shape of the escape part is relatively insensitive to the value of the potential
parameter $W_0$ in this case.

We also show the calculated results of the escape part with different
potential parameters in Figs.~\ref{fig:V70_W_p500_w_data}, ~\ref{fig:V50_W_p500_w_data}
 and ~\ref{fig:V30_W_p500_w_data} for different strengths of the
 attractive potential.
We find that the shape of the escape part is not very
sensitive to the $V_0$ and $W_0$ parameters and that it could be uneasy to
obtain the detail information on $\eta$-nucleus potential only from the
escape parts.
Although we have some cases which could be safely ruled out 
by these comparison, such as $(V_0, W_0) = -(70,5)$ and $-(50,5)$~MeV
cases, 
in which we find a distinct threshold structure 
as shown in Figs.~\ref{fig:V70_W_p500_w_data} and~\ref{fig:V50_W_p500_w_data},
the whole shape of
the calculated results are not so much different from
that of experimental data in many cases.

\begin{figure}
\begin{center}
\resizebox{0.95\hsize}{!}{%
  \includegraphics{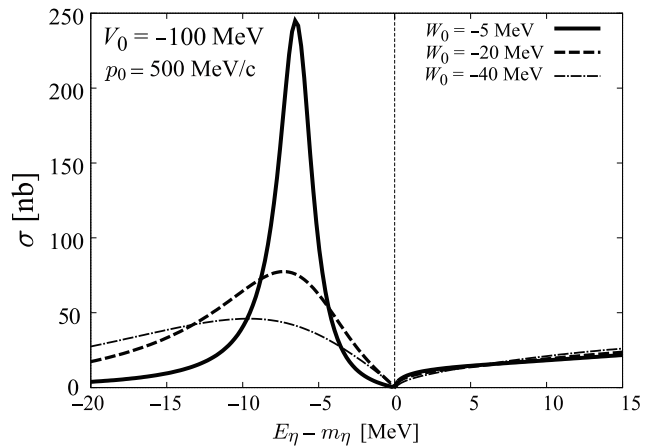} }
\caption{ Calculated total cross sections of 
$d+d \rightarrow (\eta + \alpha) \rightarrow X$ reaction 
scaled by the same factor used in Fig.~\ref{fig:V100_W_p500_w_data}
plotted as functions  
of the $\eta$ excited energy $E_\eta - m_\eta$. 
The flat contributions are subtracted. 
The parameters of the
 $\eta$-$\alpha$ optical potential are $(V_0, W_0) = -(100,5), -(100,20)$,
 and $-(100,40)$ MeV, and
 the parameter $p_0$ is fixed to be $p_0 = 500$ MeV/$c$. }
\label{fig:total_bg_V100_W_p500}       
\end{center}
\end{figure}

\begin{figure}
\begin{center}
\resizebox{0.95\hsize}{!}{%
  \includegraphics{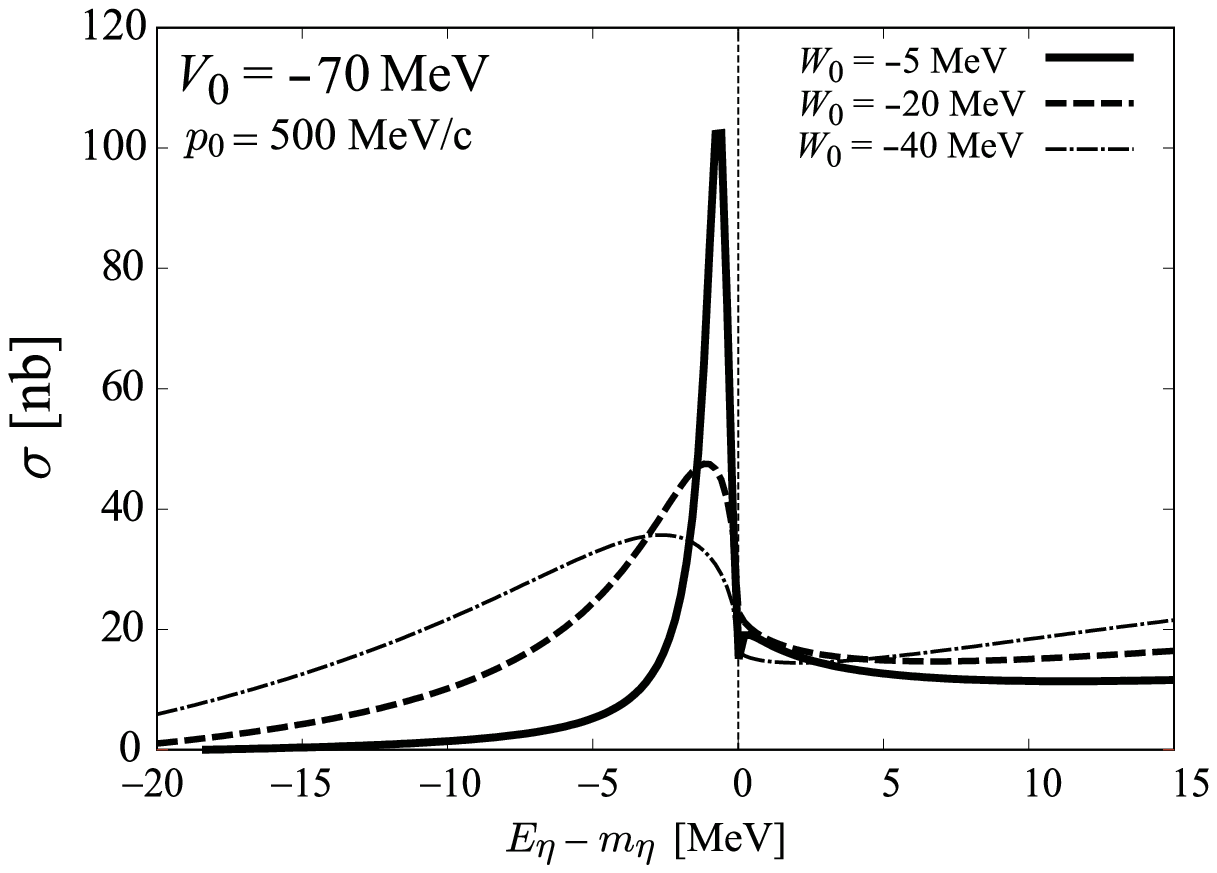}}
\caption{ Calculated total cross sections of 
$d+d \rightarrow (\eta + \alpha) \rightarrow X$ reaction 
scaled by the same factor used in Fig.~\ref{fig:V70_W_p500_w_data}
plotted as functions 
of the $\eta$ excited energy $E_\eta - m_\eta$. 
The flat contributions are subtracted. 
The parameters of the
 $\eta$-$\alpha$ optical potential are $(V_0, W_0) = -(70,5), -(70,20)$,
 and $-(70,40)$ MeV, and
 the parameter $p_0$ is fixed to be $p_0 = 500$ MeV/$c$. }
\label{fig:total_bg_V70_W_p500}       
\end{center}
\end{figure}

\begin{figure}
\begin{center}
\resizebox{0.95\hsize}{!}{%
  \includegraphics{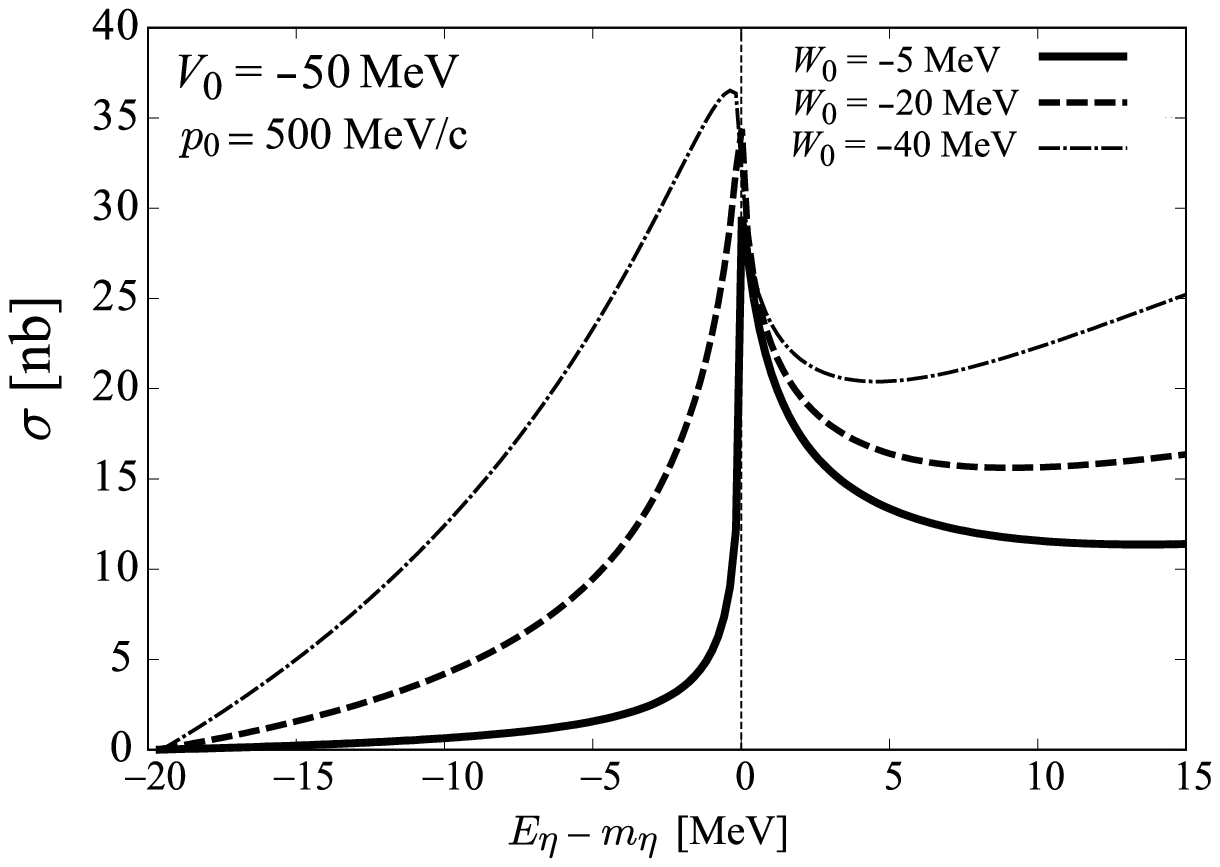}}
\caption{ Calculated total cross sections of 
$d+d \rightarrow (\eta + \alpha) \rightarrow X$ reaction 
scaled by the same factor used in Fig.~\ref{fig:V50_W_p500_w_data}
plotted as functions 
of the $\eta$ excited energy $E_\eta - m_\eta$. 
The flat contributions are subtracted. 
The parameters of the
 $\eta$-$\alpha$ optical potential are $(V_0, W_0) = -(50,5), -(50,20)$,
 and $-(50,40)$ MeV, and
 the parameter $p_0$ is fixed to be $p_0 = 500$ MeV/$c$. }
\label{fig:total_bg_V50_W_p500}       
\end{center}
\end{figure}

\begin{figure}
\begin{center}
\resizebox{0.95\hsize}{!}{%
  \includegraphics{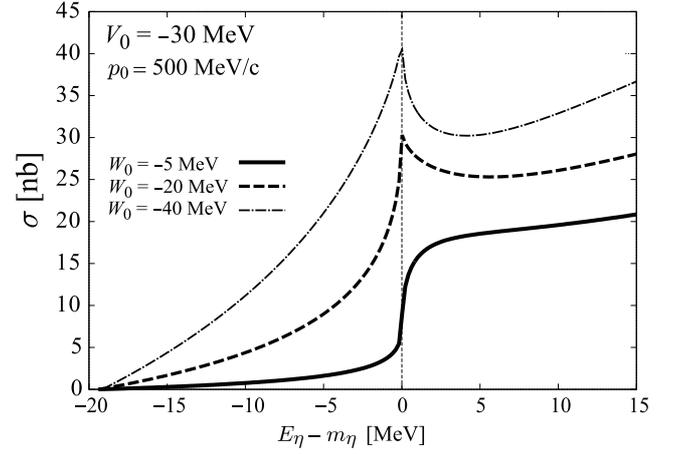}}
\caption{ Calculated total cross sections of 
$d+d \rightarrow (\eta + \alpha) \rightarrow X$ reaction 
scaled by the same factor used in Fig.~\ref{fig:V30_W_p500_w_data}
plotted as functions 
of the $\eta$ excited energy $E_\eta - m_\eta$. 
The flat contributions are subtracted. 
The parameters of the
 $\eta$-$\alpha$ optical potential are $(V_0, W_0) = -(30,5), -(30,20)$,
 and $-(30,40)$ MeV, and
 the parameter $p_0$ is fixed to be $p_0 = 500$ MeV/$c$. }
\label{fig:total_bg_V30_W_p500}       
\end{center}
\end{figure}

One of the best ways to obtain the decisive information on the
$\eta$-nucleus interaction may be to have the direct experimental 
observation of the $\eta$-nucleus spectrum below the threshold,
where the spectrum shape is more sensitive to the potential 
parameters as shown in Figs.~\ref{fig:total_V100_W_p500}--\ref{fig:total_V30_W_p500}, 
especially 
if a bound state exists. 
Nevertheless, it could happen that nature might 
not 
give us 
any bound states or our experimental techniques would not
be sufficient to distinguish a less prominent peak due to
a large width. In such cases, we could deduce the information 
on the $\eta$-nucleus interaction from the absolute value of the
spectrum 
below the $\eta$-nucleus threshold 
in comparison with the $\eta$ production cross section 
in the same reaction 
above the threshold. 

Here we have a model which 
can be used to 
calculate both the conversion 
and escape parts in the same footing simultaneously. 
The conversion part describes the spectrum shape induced by the
$\eta$ absorption to the nucleus, 
while the escape part shows the
$\eta$ production cross section. In our model, we leave 
the interaction strength of the $d+d \rightarrow \eta + \alpha$ given in 
Eq.~(\ref{eq:intH}) 
as a free parameter. We can adjust this parameter so as to 
reproduce the $\eta$ production 
data 
by the escape part of our 
spectrum, and then the conversion part can be a outcome 
from the model. Since the conversion part of the spectrum
below the threshold is more sensitive to the 
$\eta$-nucleus 
interaction 
parameters, comparing the theoretical prediction and 
experimental data of the $d+d$ reaction below the threshold,
we 
can 
deduce the information on the interaction parameters.

\begin{figure}
\begin{center}
\resizebox{0.95\hsize}{!}{%
  \includegraphics{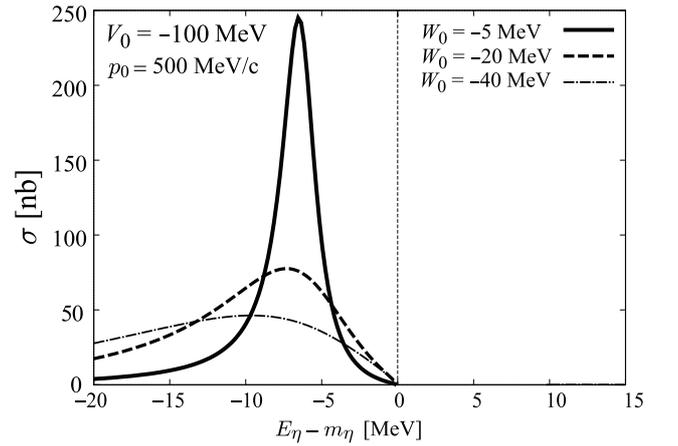} }
\caption{ Calculated conversion part of cross sections of 
$d+d \rightarrow (\eta + \alpha) \rightarrow X$ reaction 
scaled by the same factor used in Fig.~\ref{fig:V100_W_p500_w_data}
plotted as functions  
of the $\eta$ excited energy $E_\eta - m_\eta$. 
The flat contributions are subtracted. 
The parameters of the
 $\eta$-$\alpha$ optical potential are $(V_0, W_0) = -(100,5), -(100,20)$,
 and $-(100,40)$ MeV, and
 the parameter $p_0$ is fixed to be $p_0 = 500$ MeV/$c$. }
\label{fig:conv_bg_V100_W_p500}       
\end{center}
\end{figure}

\begin{figure}
\begin{center}
\resizebox{0.95\hsize}{!}{%
  \includegraphics{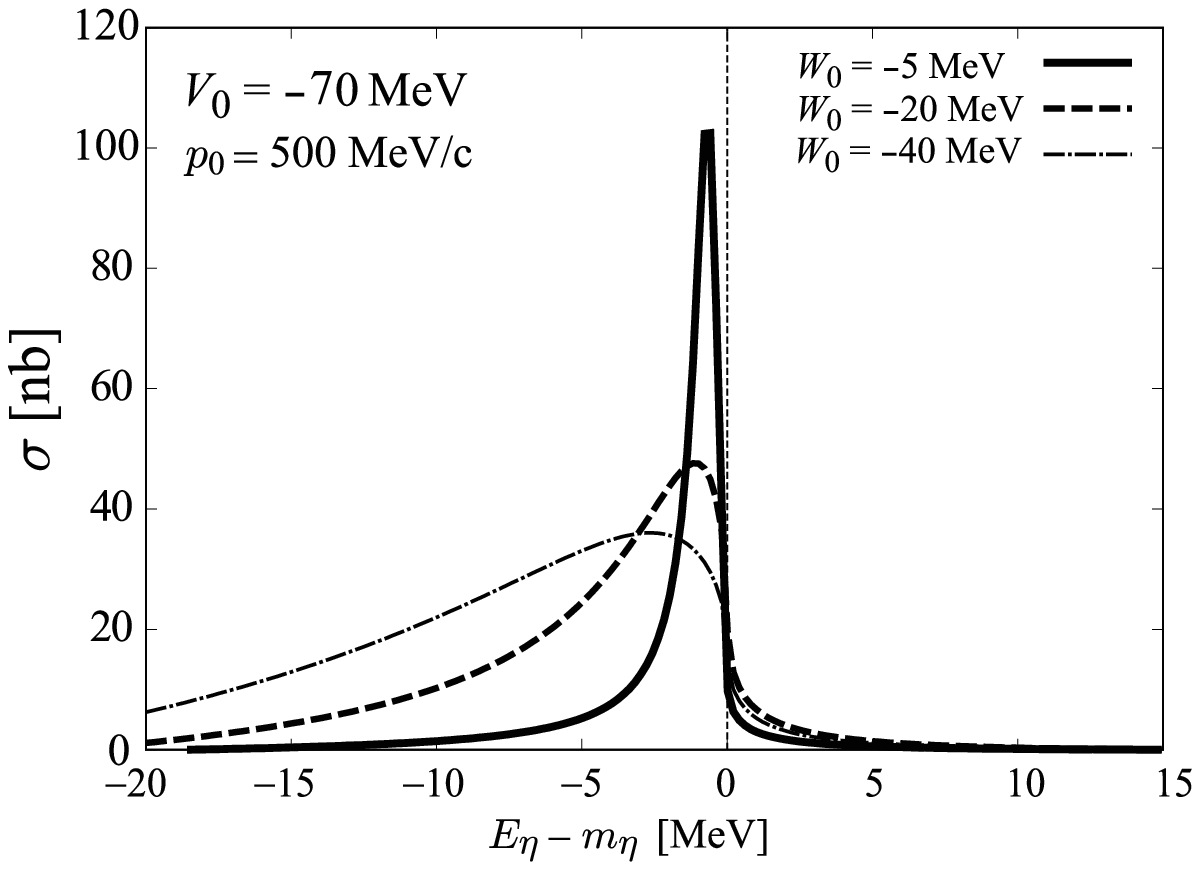}}
\caption{ Calculated conversion part of cross sections of 
$d+d \rightarrow (\eta + \alpha) \rightarrow X$ reaction 
scaled by the same factor used in Fig.~\ref{fig:V70_W_p500_w_data}
plotted as functions 
of the $\eta$ excited energy $E_\eta - m_\eta$. 
The flat contributions are subtracted. 
The parameters of the
 $\eta$-$\alpha$ optical potential are $(V_0, W_0) = -(70,5), -(70,20)$,
 and $-(70,40)$ MeV, and
 the parameter $p_0$ is fixed to be $p_0 = 500$ MeV/$c$. }
\label{fig:conv_bg_V70_W_p500}       
\end{center}
\end{figure}

\begin{figure}
\begin{center}
\resizebox{0.95\hsize}{!}{%
  \includegraphics{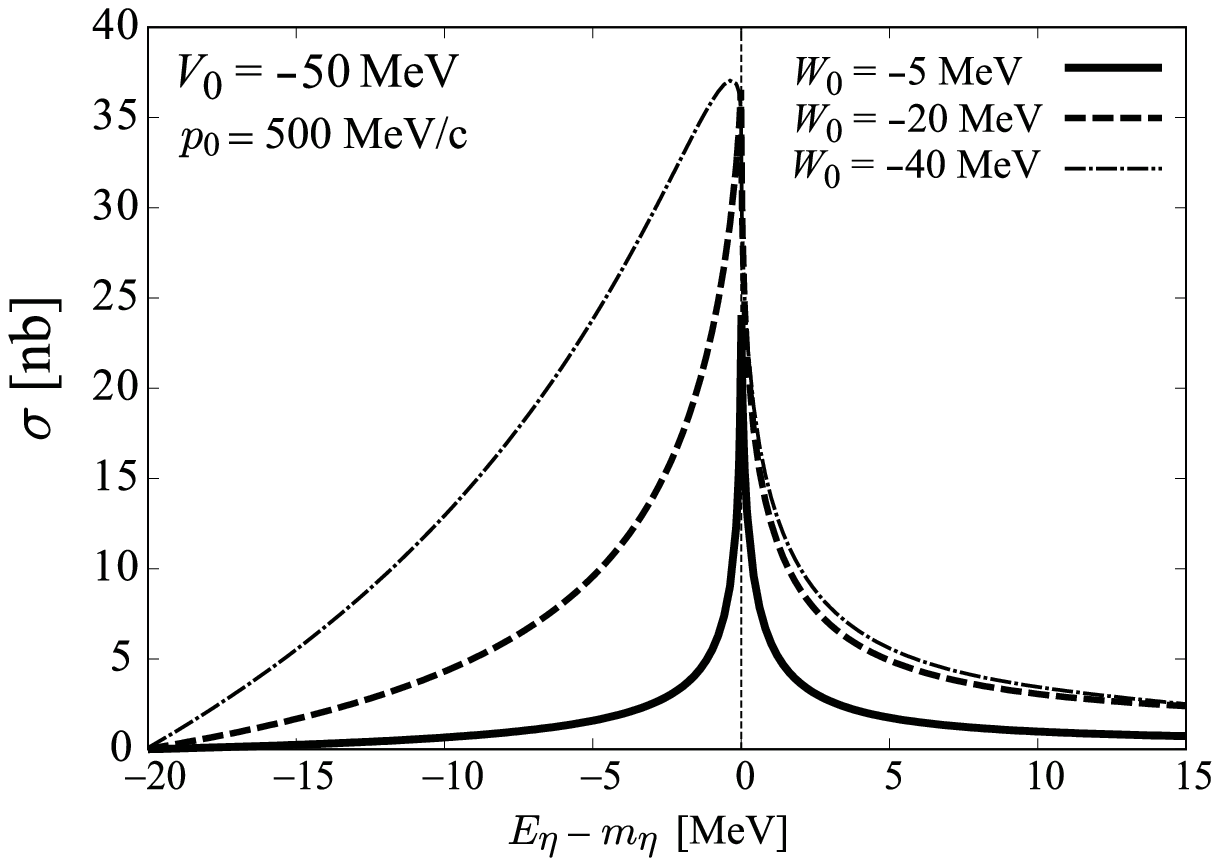}}
\caption{ Calculated conversion part of cross sections of 
$d+d \rightarrow (\eta + \alpha) \rightarrow X$ reaction 
scaled by the same factor used in Fig.~\ref{fig:V50_W_p500_w_data}
plotted as functions 
of the $\eta$ excited energy $E_\eta - m_\eta$. 
The flat contributions are subtracted. 
The parameters of the
 $\eta$-$\alpha$ optical potential are $(V_0, W_0) = -(50,5), -(50,20)$,
 and $-(50,40)$ MeV, and
 the parameter $p_0$ is fixed to be $p_0 = 500$ MeV/$c$. }
\label{fig:conv_bg_V50_W_p500}       
\end{center}
\end{figure}

\begin{figure}
\begin{center}
\resizebox{0.95\hsize}{!}{%
  \includegraphics{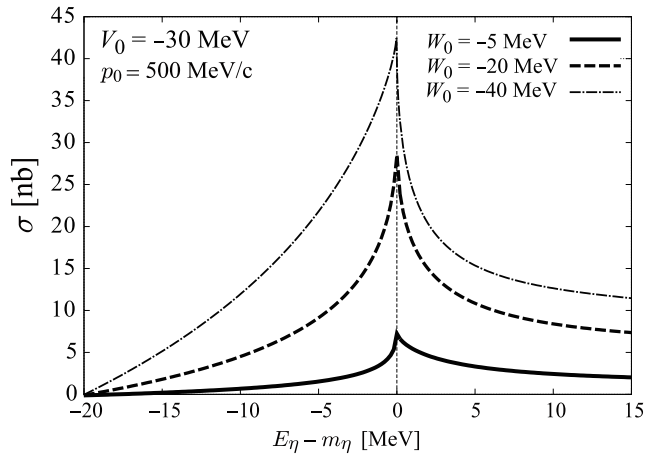}}
\caption{ Calculated conversion part of cross sections of 
$d+d \rightarrow (\eta + \alpha) \rightarrow X$ reaction 
scaled by the same factor used in Fig.~\ref{fig:V30_W_p500_w_data}
plotted as functions 
of the $\eta$ excited energy $E_\eta - m_\eta$. 
The flat contributions are subtracted. 
The parameters of the
 $\eta$-$\alpha$ optical potential are $(V_0, W_0) = -(30,5), -(30,20)$,
 and $-(30,40)$ MeV, and
 the parameter $p_0$ is fixed to be $p_0 = 500$ MeV/$c$. }
\label{fig:conv_bg_V30_W_p500}       
\end{center}
\end{figure}

For the purpose, we show the scaled theoretical total cross section of the
$d + d \rightarrow (\eta + \alpha) \rightarrow X$ reaction in 
Figs.~\ref{fig:total_bg_V100_W_p500}--\ref{fig:total_bg_V30_W_p500}
for various values of the parameter ($V_0, W_0$).
The absolute value of the cross section in these figures
are determined so that the escape part of the cross section
reproduces the experimental data of $d+d \rightarrow \eta + \alpha$ 
and the nonstructural flat contributions described above
are subtracted in the conversion part. 
%
We should mention here that the structure of the spectra in these figures
is enhanced for smaller $|W_0|$ values in 
Figs.~\ref{fig:total_bg_V100_W_p500} and \ref{fig:total_bg_V70_W_p500},
while it is suppressed for smaller $|W_0|$ values in Figs.~\ref{fig:total_bg_V50_W_p500}
and \ref{fig:total_bg_V30_W_p500}.
This behavior can be 
understood by considering the origin of the structure of the spectrum. 
For the strong attractive potential cases,
since the structure is dominated by the peak due to the existence of the bound state,
the peak structure becomes more prominent for the weaker imaginary
because of the smaller width of the bound state.
On the other hand, for the weak real potential cases,
since the structure is dominated by the absorptive processes,
the structure can be suppressed for the weaker absorptive potentials.

For comparison of our calculated results with experimental data
for the subthreshold energies,
we show only the conversion part, since in experiment 
the system energy is measured by observing a pion, nucleon 
and a residual nucleus emitted due to $\eta$ absorption 
and this process is counted in the conversion part in the 
calculation.
In Figs.~\ref{fig:conv_bg_V100_W_p500}--\ref{fig:conv_bg_V30_W_p500},
we show the calculated conversion parts of the spectra which correspond to the
$\eta$ absorption processes.
The obtained spectra shown in
Figs.~\ref{fig:conv_bg_V100_W_p500}--\ref{fig:conv_bg_V30_W_p500}
can be compared to the shape of the experimental spectra on the 
background reported in 
Ref.~\cite{Adlarson:2016dme},
where the upper limit of the peak structure of the 
$d + d \rightarrow  {^3 \rm He} + n + \pi^{0}$
 is 3--6~nb.
This implies that
the experimental upper limit for the semi-inclusive conversion 
spectrum of $d+d \rightarrow (\eta + \alpha) \rightarrow {\rm ^3He} + N+
\pi$ 
including both $n + \pi^0$ and $p + \pi^{-} $ channel
could be 
estimated to be 
3--6~nb $\times 3=$ 9--18~nb because of the isospin symmetry of the decay
channel of the $\eta$-$\alpha$ system.
In this case, the peak structures of the bound states in 
Figs.~\ref{fig:conv_bg_V100_W_p500} and~\ref{fig:conv_bg_V70_W_p500}
are fully rejected and strong attractive with less absorptive potentials are not
allowed.
%
In addition,
the upper limit provides the strong restriction to the
$\eta$-$\alpha$
potential and only weak potential cases with 
small $|V_0|$ and $|W_0|$ values 
such as $(V_0, W_0)= -(50, 5)$ and $-(30, 5)$ cases
could be allowed by the limit.

\begin{figure}
\begin{center}
 \resizebox{0.95\hsize}{!}{%
  \includegraphics{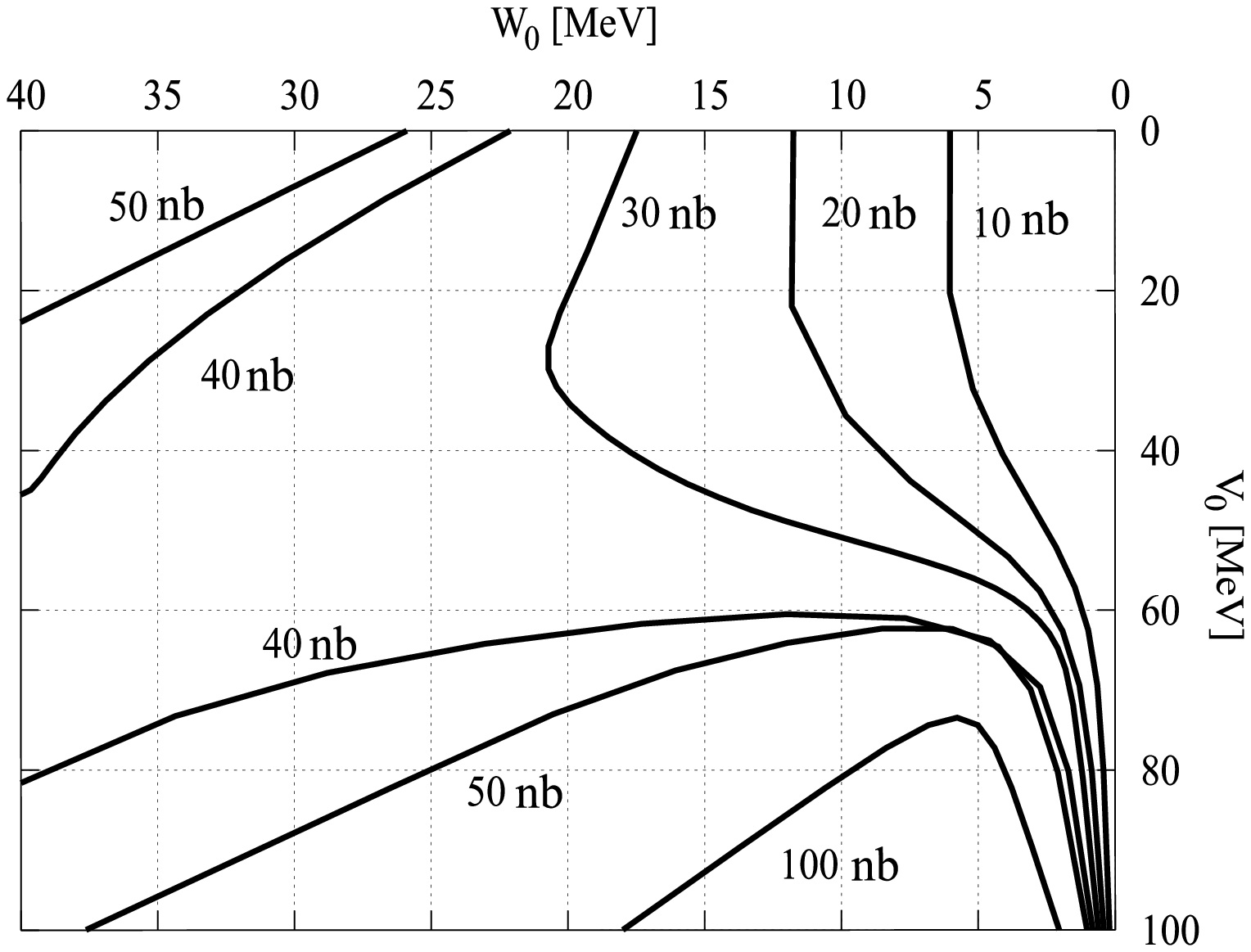}}
\caption{ 
Contour plot of the height of the structure appearing
in the conversion spectra such as shown in
Figs.~\ref{fig:conv_bg_V100_W_p500}--\ref{fig:conv_bg_V30_W_p500}
in the $V_0 - W_0$ plane.
This plot shows clearly the acceptable region
of the potential strength for various values of the upper limit of the
 height of the structure in the $d + d \rightarrow {^3 \rm He} + N + \pi$
reaction.
}
\label{fig:contour}       
\end{center}
\end{figure}

In order to understand the meaning of the experimental upper limit and
the results of our analyses more clearly, we plot a contour plot of
the height of the structure appeared in the conversion spectra on the
flat contribution in the $V_0 - W_0$
plane in Fig.~\ref{fig:contour},
where the acceptable region of $V_0$, $W_0$ values can be easily
understood for each value of the upper limit of the height of the
structure in the conversion spectra
such as shown 
in Figs.~\ref{fig:conv_bg_V100_W_p500}--\ref{fig:conv_bg_V30_W_p500}.
From this figure, the upper limit reported in
Ref.~\cite{Adlarson:2016dme} is found to provide very valuable
information on $\eta$-nucleus interaction
and strongly suggests the small $|V_0|$ and  $|W_0|$ values.
However, it should be noted that the results in Fig.~\ref{fig:contour} are
considered to be qualitative
since 
we have not considered the experimental energy resolution here.
In addition, the shapes of the structure appearing in 
Figs.~\ref{fig:conv_bg_V100_W_p500}--\ref{fig:conv_bg_V30_W_p500}
are not simply the symmetric peak.
We also need to understand the origin of the absorptive potential and
branching ratio of various decay processes to compare our results to the
data for the specific decay mode.
Actually, we have only considered one-nucleon absorption process for
$\eta$ meson here.
Multi-nucleon processes are also possible in reality~\cite{Kulpa:1998vj}.
Thus, for deducing the quantitative information on $\eta$-nucleus interaction,
it is mandatory to make the detail comparison between 
the calculated results and data
by taking account of the realistic experimental energy resolution,
asymmetric shape of the structure appeared in the conversion spectra,
the branching ratio of the decay process of $\eta$ bound states and so on,
especially for subthreshold energy region, where the spectra may have
variety of structures depending on the $\eta$-nucleus interaction
strength.

\begin{figure}
\begin{center}
\resizebox{0.95\hsize}{!}{%
 \includegraphics{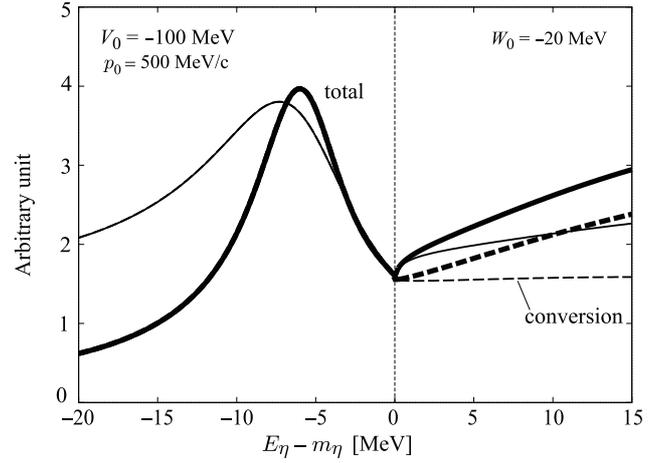}
}
\caption{Calculated cross sections of the $d + d \rightarrow (\eta +
 \alpha) \rightarrow X$ reaction for the formation of the $\eta - \alpha$ bound
 system plotted as
 functions of the $\eta$ excited energy $E_\eta - m_\eta$. The parameters of the
 $\eta$-$\alpha$ optical potential is $(V_0, W_0) = -(100,20)$ MeV, and
 the parameter $p_0$ is fixed to be $p_0 = 500$ MeV/$c$. 
The solid lines indicate the total cross
 sections $\sigma$ and the dashed lines the conversion parts
 $\sigma_{\rm conv}$.
The thin lines are the results with the energy independent optical
 potential and the thick lines are those with the energy dependent
 optical potential (see text).
}
\label{fig:total_V100_Edepend}       
\end{center}
\end{figure}

\begin{figure}
\begin{center}
\resizebox{0.95\hsize}{!}{%
 \includegraphics{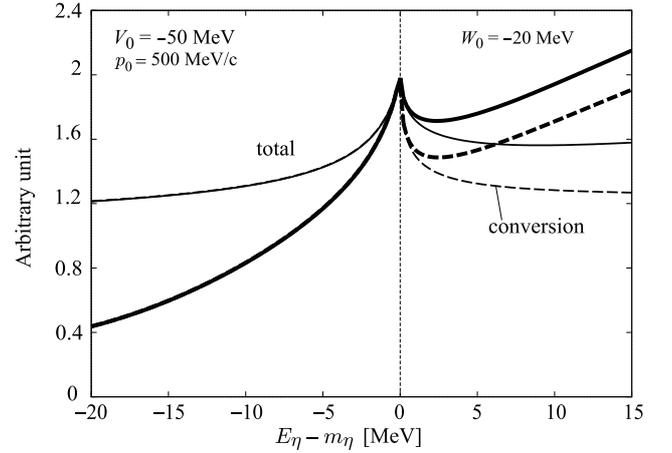}
}
\caption{Same as Fig.~\ref{fig:total_V100_Edepend} except for $(V_0, W_0) =
  -(50,20)$ MeV.}
 \label{fig:total_V50_Edepend}       
\end{center}
\end{figure}

Finally, we mention the effects of the possible energy dependence of the
optical potential. The optical potential has the energy dependence in
general and the dependence could change the calculated spectra.
To simulate the energy dependence of $\eta$-$\alpha$ optical potential, we
adopted the energy dependence of the $\eta$-nucleon scattering length in 
Ref.~\cite{Cieply:2013sya} and 
we assumed the $\eta$-nucleon relative energy 
to be the quarter of that of the $\eta$-$\alpha$.
The potential strength are normalized at the threshold energy.
The calculated results are shown in Figs.~\ref{fig:total_V100_Edepend}
and \ref{fig:total_V50_Edepend}.
We have found that the energy dependence of the imaginary part of the
optical potential mainly affects the strength of the conversion part in the
spectra and changes the flat contribution of conversion part to the
slope with some gradient.
Hence, this effect could be important for the more realistic analyses.
Though, there are many theoretical models for $\eta$-nucleon scattering
length as compiled in Ref.~\cite{Cieply:2013sga},
the qualitative features seems common for all models.

\section{Conclusion}\label{concl}
We have developed a theoretical model to evaluate the formation rate of
the $\eta$-$\alpha$ bound states in the $d + d$ fusion reaction.
Because of the difficulties due to the large momentum transfer which is
unavoidable to produce $\eta$ meson in the fusion reaction, we formulate
the model in a phenomenological way. 
We have shown the numerical results for the cases with the various sets
of the $\eta$-nucleus interaction parameters.

We have found that the data of $\eta$ production above threshold provide
important information on the absolute strength of the reaction 
by comparing them with the escape part of the calculated results.
The upper limit of the formation cross section of the $\eta$ mesic
nucleus reported in Ref.~\cite{Skurzok:2016fuv} also provides the significant
information on the strength of the $\eta$-nucleus interaction.
We would like to stress here that 
simultaneous fit to both data of $d+d \rightarrow \eta + \alpha$ and 
$d +d \rightarrow (\eta + \alpha) \rightarrow X$ 
using our model make it possible to provide  
valuable information on $\eta$-nucleus interaction.
The results of our analyses are compiled in Fig.~\ref{fig:contour} as
a contour plot of the $V_0 - W_0$ plane.


The present discussion is simply based on the value of the
upper limit of the peak structure in the fusion reaction spectrum
below the threshold. As for the further works, to make the analyses
performed in this article more quantitative and developed,
direct comparison of the spectrum shapes between the calculated
results and experiments should be necessary. For this purpose,
we should take account of experimental energy resolution in the 
calculation and consider other possibilities of the shapes of 
the spectrum structure by improving the $\eta$-nucleus optical 
potential.

\begin{acknowledgement} 
We acknowledge the fruitful discussion with 
P. Moskal, W. Krzemien, and M. Skurzok. 
S.~H. thanks A.~Gal, N.~G.~Kelkar, S.~Wycech, E.~Oset and V.~Metag for fruitful comments
 and discussion in Krakow.
We also thank K.~Itahashi and H.~Fujioka for many discussions and
 collaborations on meson-nucleus systems.
This work was partly supported by 
JSPS KAKENHI Grant Numbers JP24540274 and JP16K05355 (S.H.), 
17K05443 (H.N.), JP15H06413 (N.I.), and JP17K05449 (D.J.)
in Japan.
\end{acknowledgement}

\appendix 
\setcounter{equation}{0} %
\renewcommand{\theequation}{\Alph{section}.\arabic{equation}}
\setcounter{figure}{0} %
\renewcommand{\thefigure}{\Alph{section}.\arabic{figure}}

\section{Appendix}\label{App}

 In this appendix, we show the numerical results for the different
 functional form of the transition form factor.
We consider the two functions defined as
\begin{equation}
 f_1 ({\vec r}) =  \left(\frac{m}{2\pi} \right)^{1/2} \frac{e^{-mr}}{r}
\label{Form:f1}
\end{equation}
with $m= p_0/\sqrt{6}$, and 
\begin{equation}
 f_2 ({\vec r}) =  \left(\frac{\lambda^3}{\pi} \right)^{1/2} e^{- \lambda r } 
\label{Form:f2}
\end{equation}
with $\lambda = p_0$ as different forms of the transition form factor. 
The Gaussian form defined in Eq.~(\ref{eq:F}) corresponds to 
\begin{equation}
 f({\vec r})   
 = \left(\frac{p_{0}^2 }{2\pi} \right)^{3/4} 
 \exp\left[ - \frac{p_{0}^2 r^2}{4}\right],
\label{Form:f}
\end{equation}
in the coordinate space.
The parameters $m$ and $\lambda$ are fixed to reproduce the same `root
mean square radius' $\displaystyle  \left( \int |f|^2 r^2  d \vec{r} \right)^{1/2}$ with the Gaussian form factor.

\begin{figure}
\begin{center}
\resizebox{0.95\hsize}{!}{%
 \includegraphics{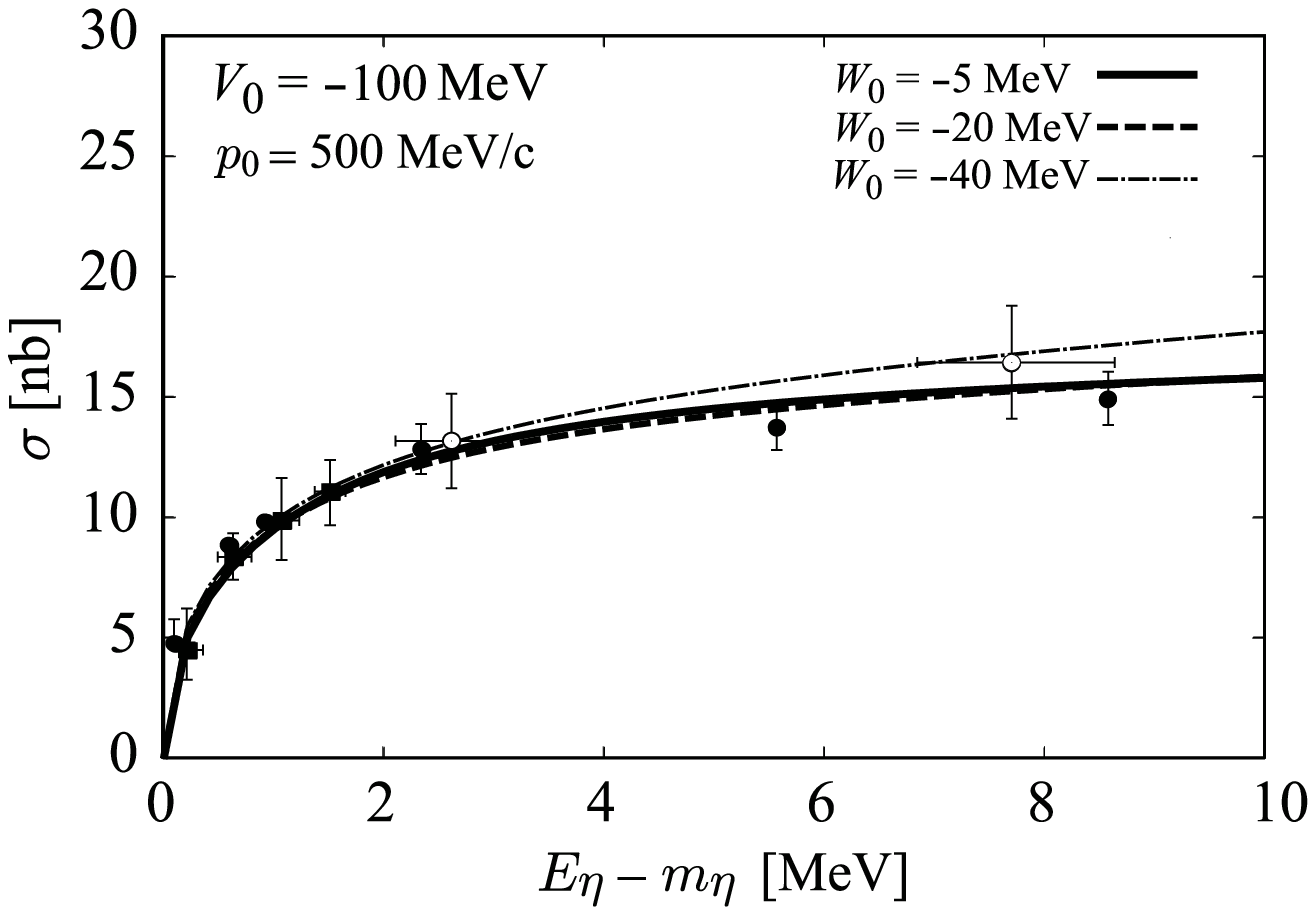} }
\caption{ 
Same as Fig.~\ref{fig:V100_W_p500_w_data} except for the transition form
 factor $f_1(\vec{r})$ defined in Eq.~(\ref{Form:f1}) is used instead of that in 
Eqs.~(\ref{eq:F}) and (\ref{Form:f}).
}
\label{fig:yukawa_V100_W_p500_w_data}       
\end{center}
\end{figure}

\begin{figure}
\begin{center}
\resizebox{0.95\hsize}{!}{%
 \includegraphics{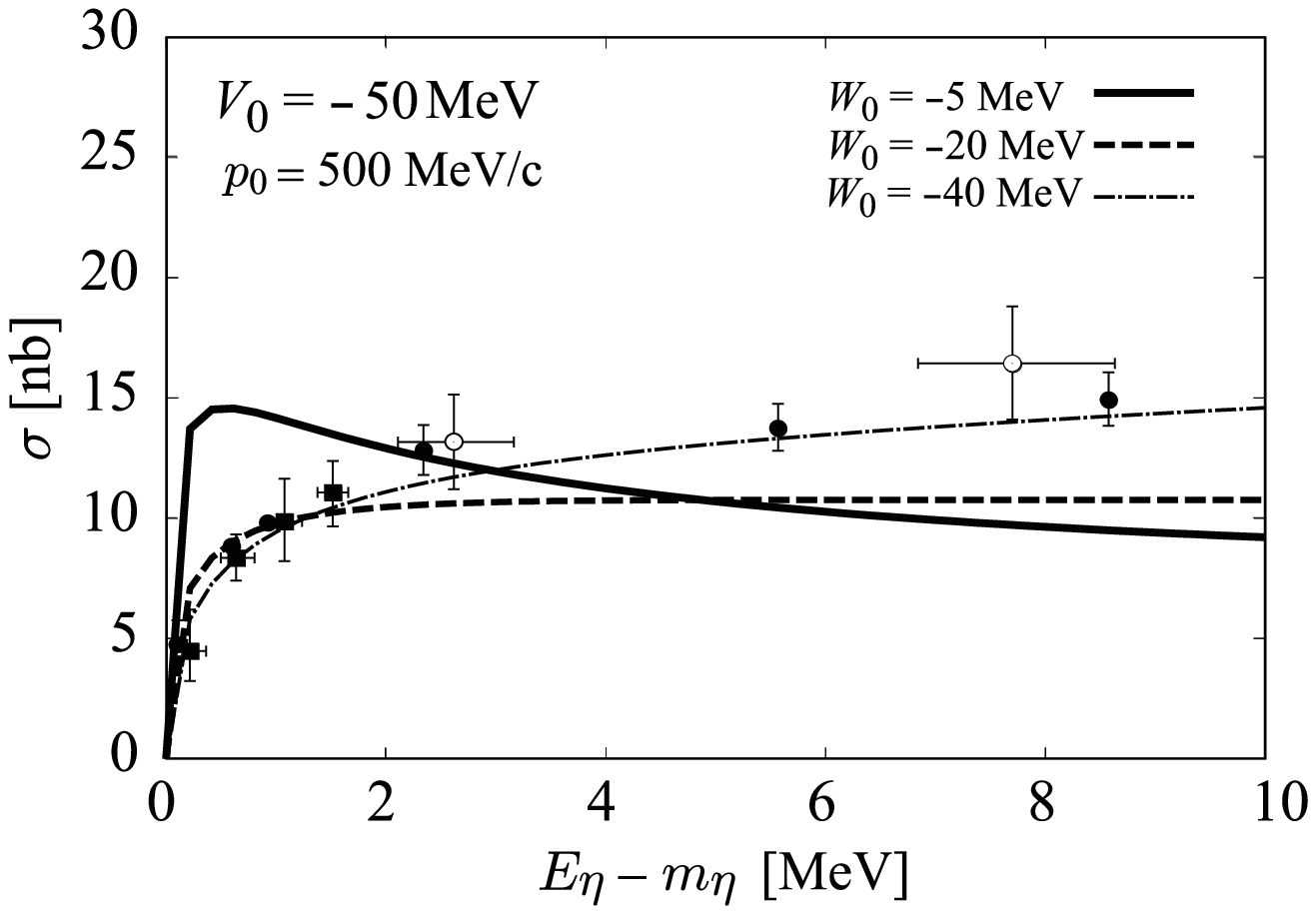} }
\caption{ 
Same as Fig.~\ref{fig:V50_W_p500_w_data} except for the transition form
 factor $f_1(\vec{r})$ defined in Eq.~(\ref{Form:f1}) is used instead of that in 
Eqs.~(\ref{eq:F}) and (\ref{Form:f}).
}
\label{fig:yukawa_V50_W_p500_w_data}       
\end{center}
\end{figure}

\begin{figure}
\begin{center}
\resizebox{0.95\hsize}{!}{%
  \includegraphics{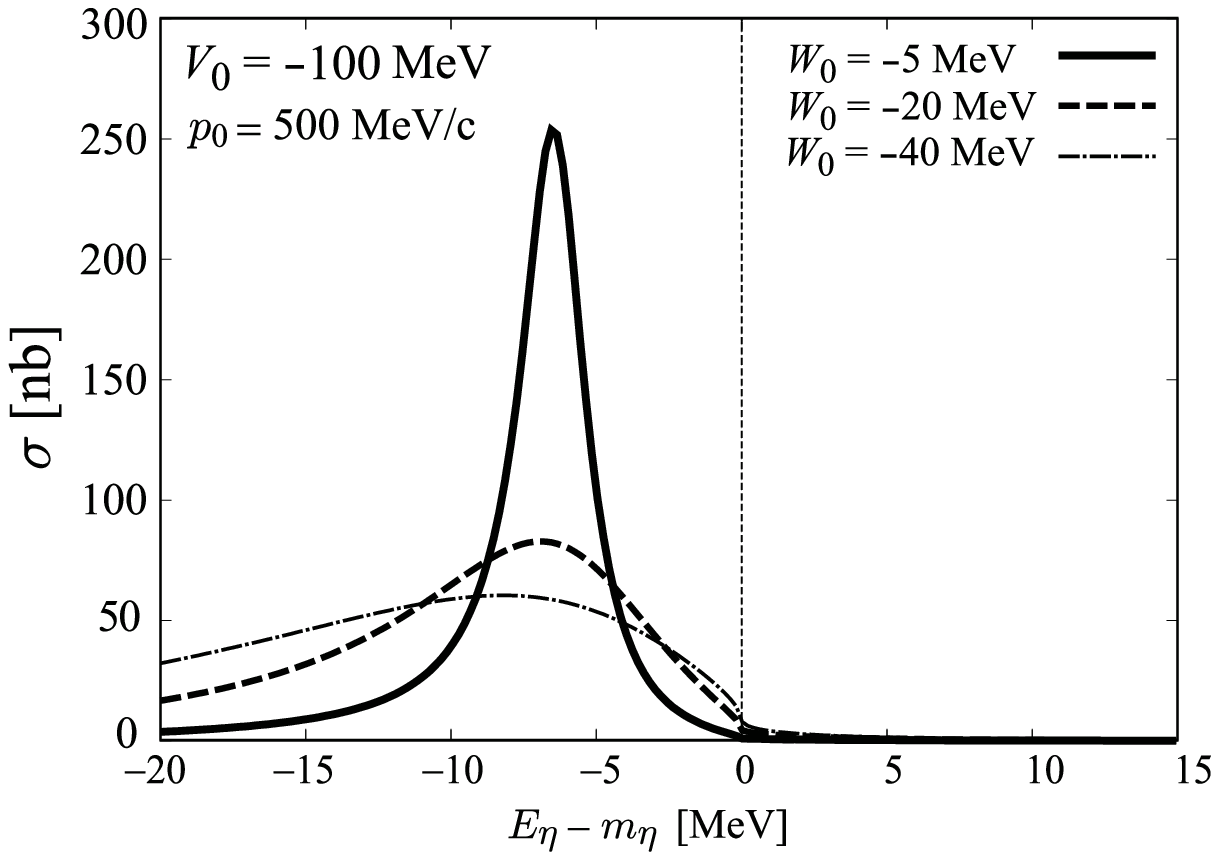} }
\caption{ 
Same as Fig.~\ref{fig:conv_bg_V100_W_p500} except for the transition form
 factor $f_1(\vec{r})$ defined in Eq.~(\ref{Form:f1}) is used instead of that in 
Eqs.~(\ref{eq:F}) and (\ref{Form:f}).
}
\label{fig:yukawa_conv_bg_V100_W_p500}       
\end{center}
\end{figure}

\begin{figure}
\begin{center}
\resizebox{0.95\hsize}{!}{%
  \includegraphics{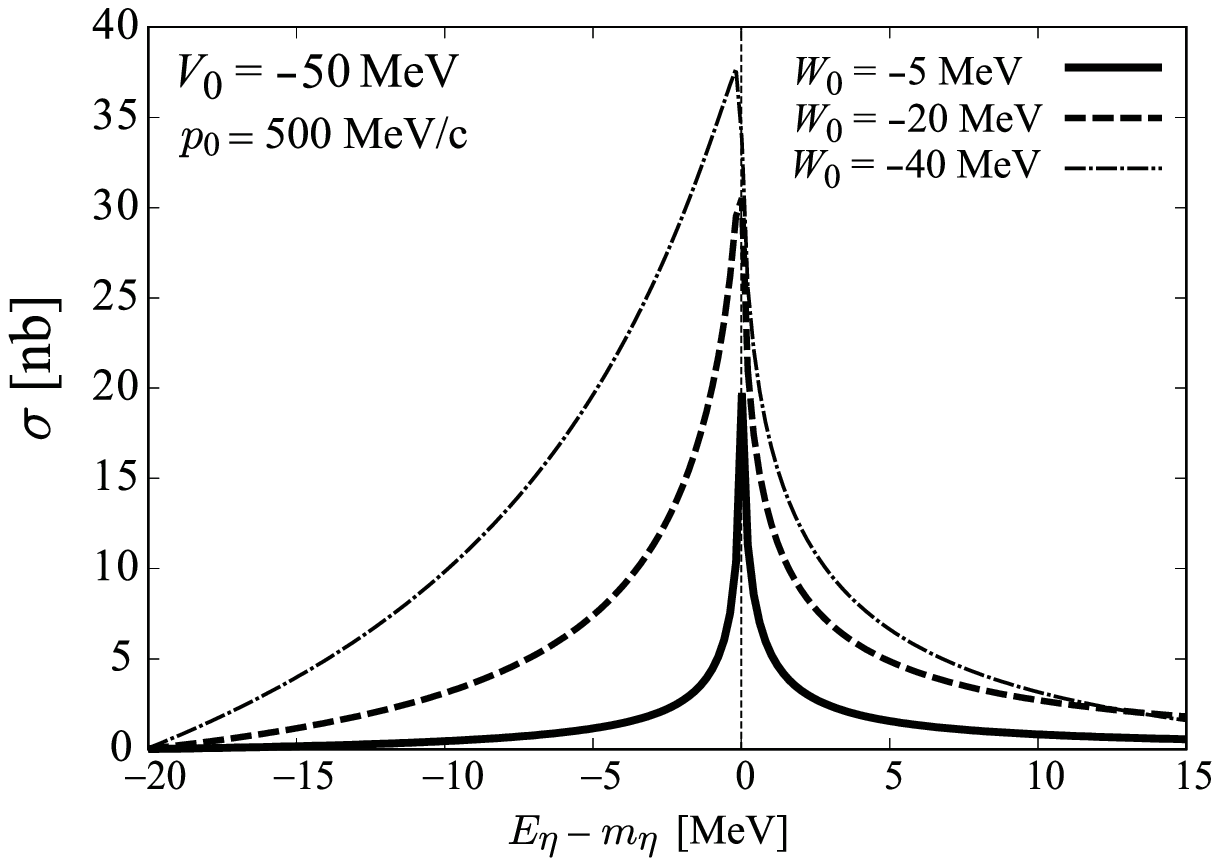} }
\caption{ 
Same as Fig.~\ref{fig:conv_bg_V50_W_p500} except for the transition form
 factor $f_1(\vec{r})$ defined in Eq.~(\ref{Form:f1}) is used instead of that in 
Eqs.~(\ref{eq:F}) and (\ref{Form:f}).
}
\label{fig:yukawa_conv_bg_V50_W_p500}       
\end{center}
\end{figure}

\begin{figure}
\begin{center}
\resizebox{0.95\hsize}{!}{%
 \includegraphics{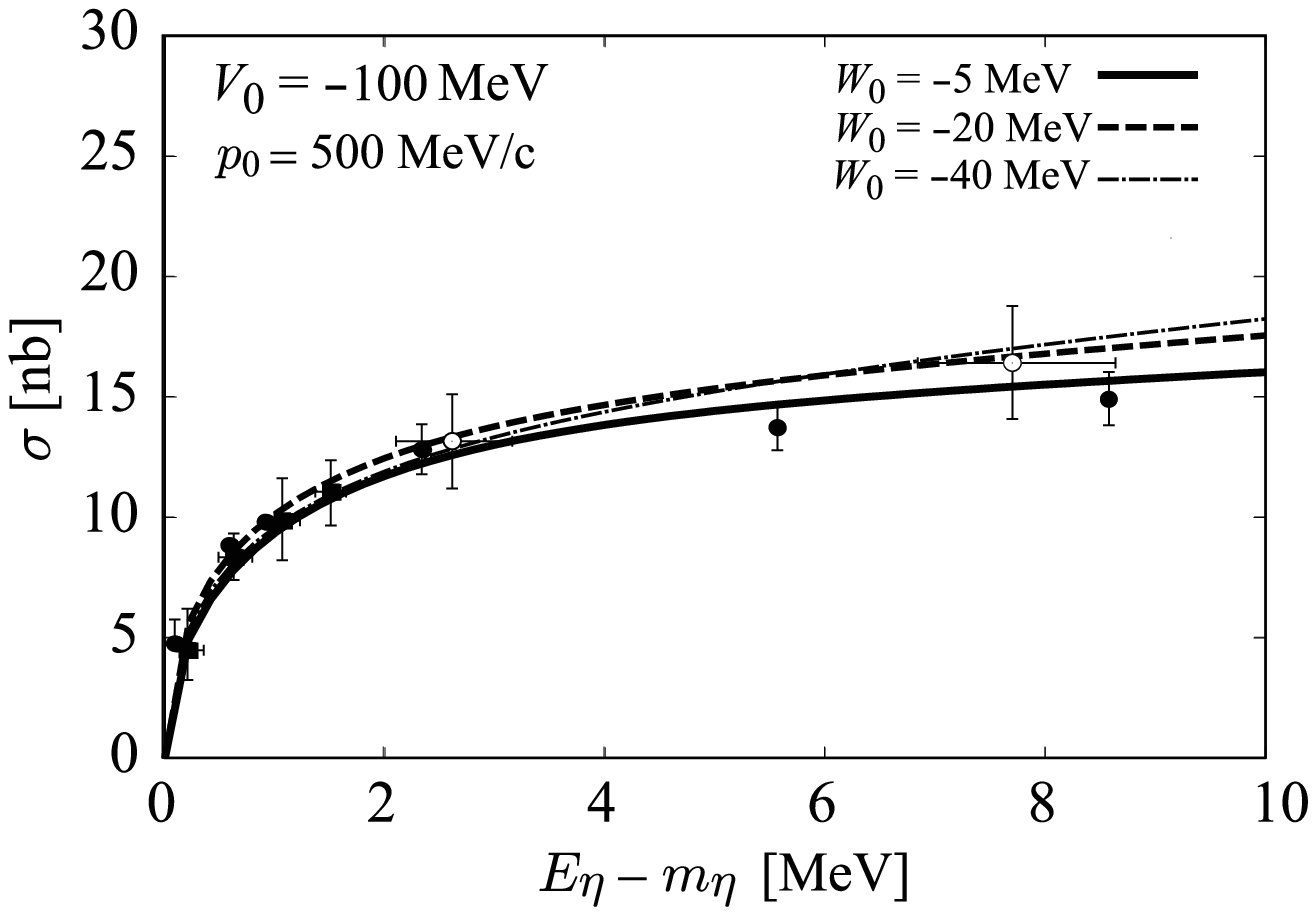} }
\caption{ 
Same as Fig.~\ref{fig:V100_W_p500_w_data} except for the transition form
 factor $f_2(\vec{r})$ defined in Eq.~(\ref{Form:f2}) is used instead of that in 
Eqs.~(\ref{eq:F}) and (\ref{Form:f}).
}
\label{fig:Exp_V100_W_p500_w_data}       
\end{center}
\end{figure}

\begin{figure}
\begin{center}
\resizebox{0.95\hsize}{!}{%
 \includegraphics{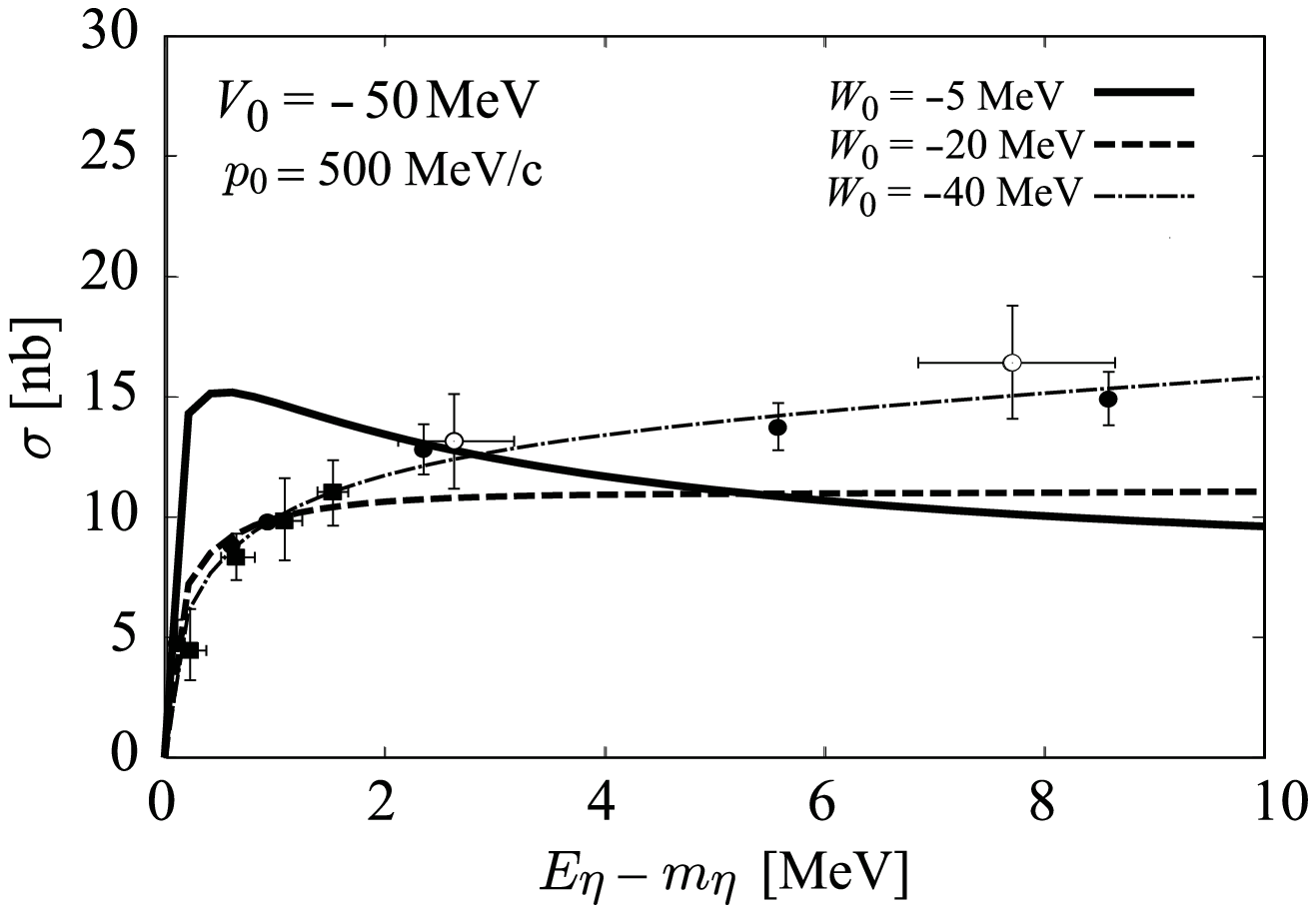} }
\caption{ 
Same as Fig.~\ref{fig:V50_W_p500_w_data} except for the transition form
 factor $f_2(\vec{r})$ defined in Eq.~(\ref{Form:f2}) is used instead of that in 
Eqs.~(\ref{eq:F}) and (\ref{Form:f}).
}
\label{fig:Exp_V50_W_p500_w_data}       
\end{center}
\end{figure}

\begin{figure}
\begin{center}
\resizebox{0.95\hsize}{!}{%
  \includegraphics{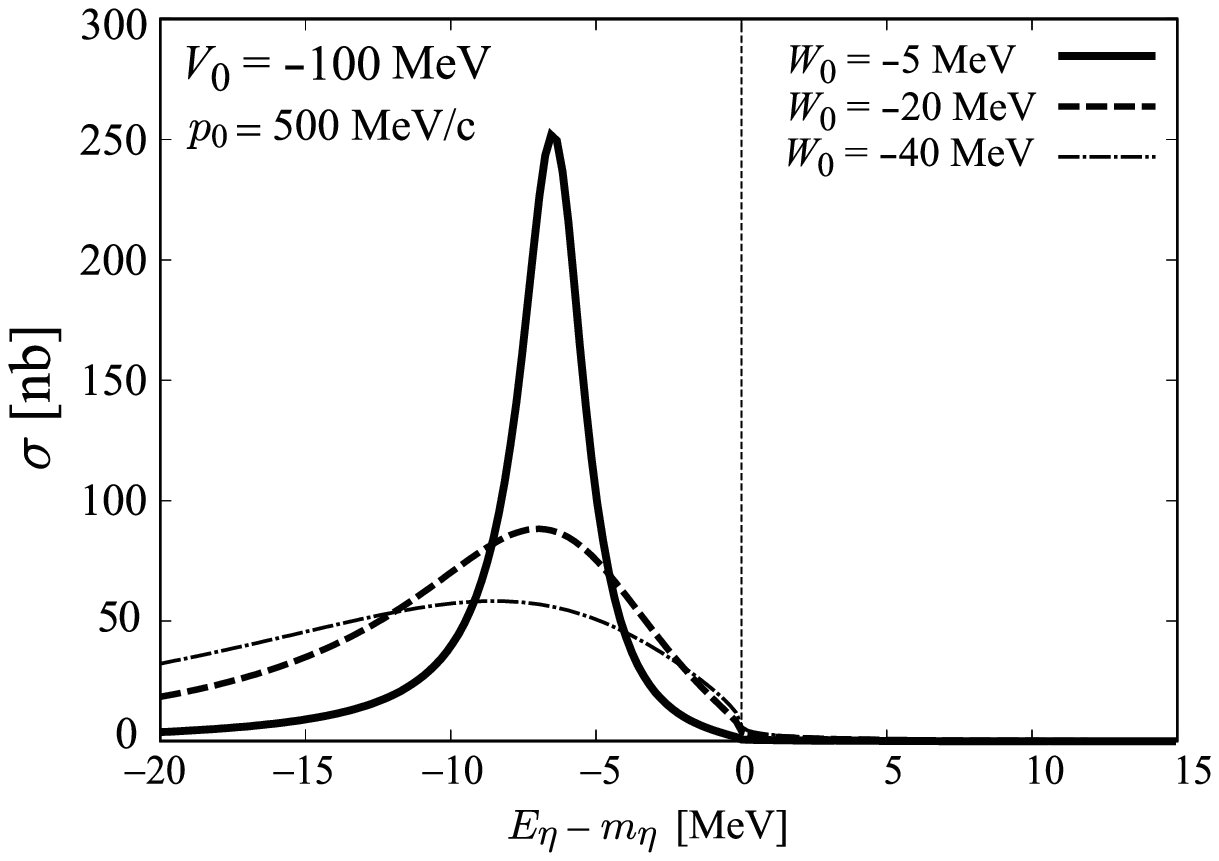} }
\caption{ 
Same as Fig.~\ref{fig:conv_bg_V100_W_p500} except for the transition form
 factor $f_2(\vec{r})$ defined in Eq.~(\ref{Form:f2}) is used instead of that in 
Eqs.~(\ref{eq:F}) and (\ref{Form:f}).
}
\label{fig:Exp_conv_bg_V100_W_p500}       
\end{center}
\end{figure}

\begin{figure}
\begin{center}
\resizebox{0.95\hsize}{!}{%
  \includegraphics{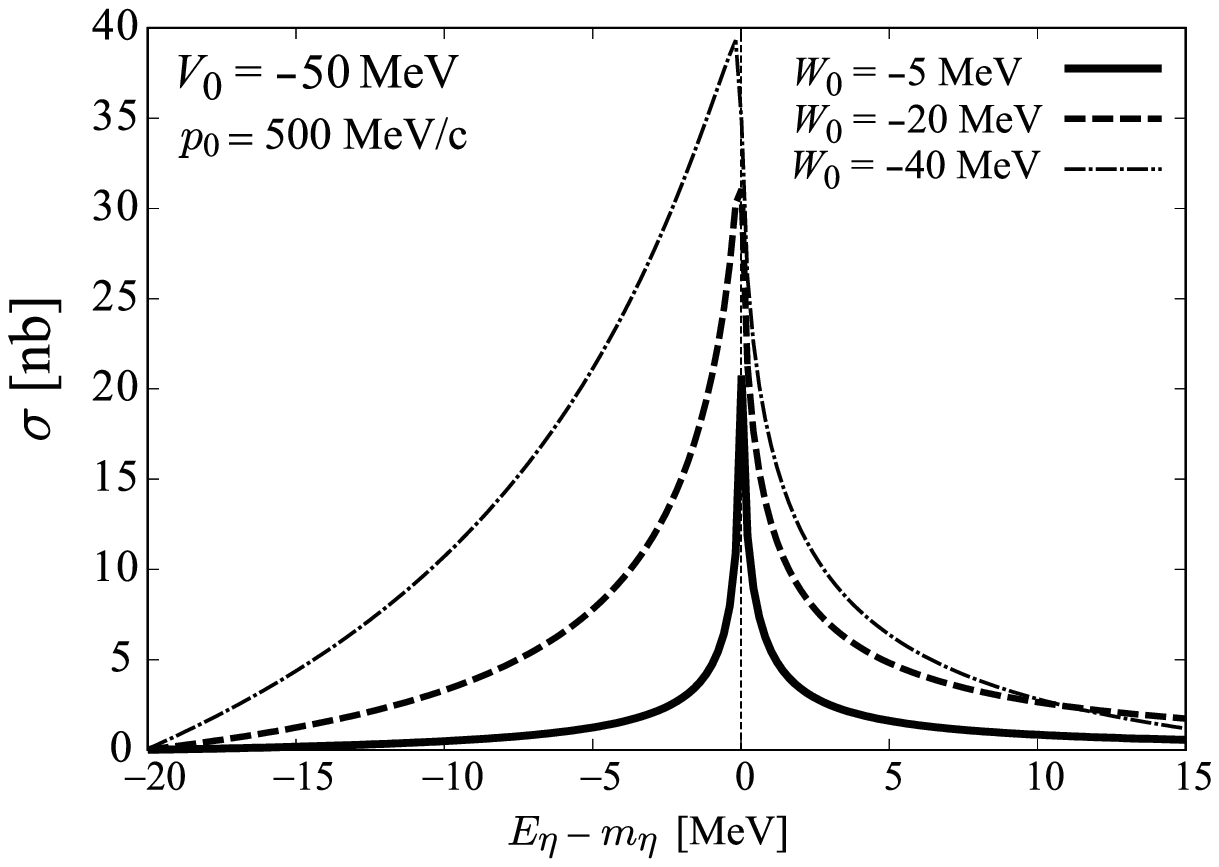} }
\caption{ 
Same as Fig.~\ref{fig:conv_bg_V50_W_p500} except for the transition form
 factor $f_2(\vec{r})$ defined in Eq.~(\ref{Form:f2}) is used instead of that in 
Eqs.~(\ref{eq:F}) and (\ref{Form:f}).
}
\label{fig:Exp_conv_bg_V50_W_p500}       
\end{center}
\end{figure}

We show the calculated results with $f_1(\vec{r})$ in
Figs.~\ref{fig:yukawa_V100_W_p500_w_data}--\ref{fig:yukawa_conv_bg_V50_W_p500}
and results with $f_2(\vec{r})$ in
Figs.~\ref{fig:Exp_V100_W_p500_w_data}--\ref{fig:Exp_conv_bg_V50_W_p500},
which correspond to Figs.~\ref{fig:V100_W_p500_w_data},
\ref{fig:V50_W_p500_w_data}, \ref{fig:conv_bg_V100_W_p500},  
\ref{fig:conv_bg_V50_W_p500}  
obtained with $f(\vec{r})$ with $p_0 =500$~MeV/$c$.
These results are observables which can be compared
with the appropriate experimental data.
We have found that all results resemble each other
and that the numerical results are robust to the choice of the
functional form of the transition form factor.

%
%
%
%
%

\end{document}